\documentclass[11pt,tightenlines,eqsecnum,floats,aps,amsmath,amssymb,nofootinbib,prd,shownopacs,floatfix]{revtex4-2}
\usepackage{graphicx}
\usepackage{epstopdf}
\usepackage{latexsym}
\usepackage{amssymb}
\usepackage{amsmath}
\usepackage{color}
\usepackage{mathrsfs}
\usepackage{xparse}
\usepackage{float}
\usepackage{mathtools}
\usepackage[normalem]{ulem}

\usepackage[center]{subfigure}

 \newcommand{\bq}{\begin{equation}}
 \newcommand{\eq}{\end{equation}}
 \newcommand{\bqn}{\begin{eqnarray}}
 \newcommand{\eqn}{\end{eqnarray}}
 
 \newcommand{\lb}{\label}

\NewDocumentCommand{\evalat}{sO{\big}mm}{%

  \IfBooleanTF{#1}
   {\mleft. #3 \mright|_{#4}}
   {#3#2|_{#4}}%
}

\begin{document}

\title{On the role of dissipative effects in the quantum gravitational onset of warm Starobinsky inflation in a closed universe}

\author{Meysam Motaharfar}
\email{mmotah4@lsu.edu}
\author{Parampreet Singh}
\email{psingh@lsu.edu}
\affiliation{Department of Physics and Astronomy, Louisiana State University, Baton Rouge, LA 70803, USA}

\begin{abstract}
A problematic feature of low energy scale inflationary models, such as Starobinsky inflation, in a spatially closed universe is the occurrence of a recollapse and a big crunch singularity before inflation can even set in. In a recent work it was shown that this problem can be successfully resolved in loop quantum cosmology for a large class of initial conditions due to a non-singular cyclic evolution and a hysteresis-like phenomena. However, for certain highly unfavorable initial conditions the onset of inflation was still difficult to obtain. In this work, we explore the role of dissipative particle production, which is typical in warm inflation scenario, in the above setting.  We find that entropy production sourced by such  dissipative effects makes hysteresis-like phenomena stronger. As a result, the onset of inflation is quick in general including for highly unfavorable initial conditions where it fails or is significantly delayed  in the absence of dissipative effects. We phenomenologically consider three warm inflation scenarios with distinct forms of dissipation coefficient, and from dynamical solutions and phase space portraits find that the phase space of favorable initial conditions turns out to be much larger than in cold inflation.

\end{abstract}

\maketitle


\section{Introduction}

Inflation is a finite period of quasi-de Sitter accelerated expansion  in the early universe which  elegantly predicts the minimal late-time curvature as well as reproduces the  adiabatic,  nearly-Gaussian, and quasiscale-invariant spectrum of primordial density fluctuations in  accordance  with  the  observational  cosmological  data. An important issue in inflationary models is that of right initial conditions for the inflaton to successfully yield sufficient e-foldings to confirm with observations. 
This issue becomes more relevant in the case of low energy inflationary models such as with Starobinsky potential \cite{Starobinsky} which are favored by observations \cite{Planck2015, Planck2018}.  Starting from Planck regime, the potential energy is suppressed in low energy inflation models and inflaton starts with kinetic energy
 domination \cite{lindelecture, linde2018}. If the universe is spatially closed then such a model can undergo a recollapse before the onset of inflation and encounter the big crunch singularity. It has been expected that a quantum theory of gravity may shed some insights on this issue. 
Since the main problem in above scenario is the existence of a recollapse followed by a big crunch singularity, if quantum gravity effects can resolve the big crunch singularity and result in a non-singular cyclic evolution then one can hope that in subsequent cycles conditions on dominance of kinetic versus the potential energy alter in such a way that recollapse can be avoided and inflation can begin. 

Before we investigate the above problem in this manuscript, it is important to make some remarks to set the right context of this study and discuss alternative strategies to solve above problem. 
Our study is based on assuming a positive spatial curvature of the universe. It has been noted earlier that one requires a high degree of fine tuning to start inflation in low energy models with a positive spatial curvature \cite{linde2003}. Thus, in a sense we take the most difficult case to understand the initial conditions problem because if the universe is spatially-flat or spatially open the recollapse caused by intrinsic curvature is absent. In fact, the initial condition problem in such cases, especially with a compact topology, becomes much easier to address \cite{compact, compact2, lindelecture}. Though it has been recently claimed that  a primordial spatial curvature may partially account for the observed anomaly in the temperature anisotropy spectrum at low multipoles \cite{low-multipoles}, and a small amount of late-time curvature consistent with current observational data has the potential to explain the current discrepancy between dataset probing early universe and those exploring late-time universe properties \cite{bharat3,DiValentino:2019qzk,handley}, when Planck results are combined with baryon acoustic oscillations data one finds that the current observations are consistent with a spatially-flat universe \cite{Planck2018, Planck20}. But the almost spatial flatness of the universe in current epoch does not imply that it was spatially flat in the pre-inflationary epoch. Thus it is worthwhile to study all the cases of spatial curvature to understand the problem of initial conditions. Let us also note that if the universe has a positive spatial curvature, the problem of recollapse can be avoided in low energy inflationary models by considering an additional field in a quadratic potential or a similar potential which drives inflation in the beginning which is carried over by the low energy inflation.\footnote{We thank the anonymous referee for pointing out this possibility.} In such a model, the additional field starts from initial conditions which are potential dominated at Planck density such that the problem of recollapse is completely avoided before the low energy inflation onsets. This strategy is expected to work for any other low energy inflation model  with an additional scalar field sourced by a potential allowing the first phase of inflation to start near Planck density. Another possibility is to consider alternatives to low energy inflation models, such as a chaotic inflationary model with additional cubic and quartic terms which turn out to be consistent with the Planck data \cite{lindelecture, linde2018}.


While above strategies exist to alleviate the problem of initial conditions in low energy inflation models, our objective in this study is to understand whether quantum gravity effects when included can resolve this problem without any additional fields which start inflation near Planck density. Since the big crunch singularity caused by a recollapse in the pre-inflationary phase is a roadblock to solve this problem, it is pertinent to incorporate quantum gravity modifications which resolve the big crunch singularity to understand the onset of inflation in low energy inflation models. 
This problem was recently addressed using non-perturbative quantum gravitational effects in loop quantum cosmology (LQC) \cite{gls20}. It was shown that although a large class of unfavorable initial conditions do not result in inflation in the classical theory and lead to a big crunch singularity in a few Planck seconds, the universe successfully goes through an inflationary phase after multiple non-singular cycles of expansion and contraction due to quantum gravity effects. The goal of the current work lies in the same direction with an aim to improve and generalize these results to demonstrate that inclusion of dissipative particle production in LQC results in a rather quick, and more robust onset of inflation even for those extreme initial conditions where inflation does not occur with above quantum gravity effects.

 Let us recall that the   non-perturbative loop quantum gravitational effects resolve the big bang/big crunch singularities replacing them by a non-singular bounce when energy density reaches Planckian values \cite{aps1,aps3,slqc}. For the spatially-closed model, singularity resolution results in multiple non-singular cycles of expansion and contraction \cite{apsv,warsaw,ck2011}. It is to be noted that loop quantum gravity effects are only dominant near the classical singularities and diminish quickly at smaller energy densities resulting in classical dynamics at the  macroscopic scales. In an effective spacetime description of these quantum gravity effects,  modified Friedmann equations can be obtained which have been shown to capture the underlying quantum dynamics to an excellent approximation \cite{aps3,apsv,numlsu-2,numlsu-4}. From these modified Friedmann equations, one can show a generic resolution of all strong curvature singularities in isotropic and anisotropic models in LQC \cite{generic} including in the presence of spatial curvature \cite{psvt}. Given that LQC robustly solves the problem of singularities, it provides an excellent stage to address the problem  of resolution of onset of inflation in low energy inflationary models in presence of a positive spatial curvature.

 An interesting feature of cosmic expansion/contraction which leads to a novel hysteresis  like phenomena in non-singular cyclic evolution is the difference in pressure during expansion and contraction stages \cite{Tolman:1939jz}. This phenomena occurs even  in the absence of dissipative effects for suitable scalar field potentials
 \cite{Sahni:2012er}. Hence, the universe posses an arrow of time due to an asymmetric equation of state during expansion-contraction phase \cite{Sahni:2015kga} rather than entropy production due to viscous pressure as it was the case in Tolman's model \cite{Tolman:1939jz}.  Of course, the most challenging issue to build such models is to overcome big bang/crunch singularities and to achieve a non-singular evolution. This task was completed in LQC where hysteresis-like phenomena was demonstrated for chaotic $\phi^2$ inflation  \cite{Dupuy:2019ibu}, a result which was recently generalized for Starobinsky inflation \cite{gls20}.  An interesting feature of such hysteresis-like period is that although the universe may fail to inflate at first, conditions improve in subsequent cycles for the onset of inflation because the ratio of kinetic to potential energy decreases and subsequent  equation of state $\omega$, defined as the ratio of total pressure and energy density, becomes less than $-1/3$. This causes a phase of accelerated expansion and as a result the recollapse is avoided. This phenomena of occurrence of non-singular cyclic evolution followed by inflation turns out to be a feature of a large class of initial conditions for $\phi^2$ and Starobinsky inflation models \cite{gls20}. However, for the latter the onset of inflation is found to require a much larger number of cycles in contrast to the $\phi^2$ inflation and for certain highly unfavorable initial conditions, inflation was not found to occur even after numerous non-singular cycles of expansion and contraction \cite{gls20}. The reason for this was tied to the weak hysteresis for low energy inflation models.

To overcome the problem of onset of inflation in such cases we note that
dissipation  is  an  indispensable  part  of  any physical system interacting with its environment, and  there are two different dynamical realizations for inflation: cold inflation and warm inflation \cite{Warm inflation}, depending on whether non-equilibrium dissipative particle production processes due to the couplings of the inflaton field with other field degrees of freedom are negligible or not during inflation. In  fact,  dissipative  processes  determine  the way ultimately the vacuum energy density,  stored in the inflaton field, ends up converting into radiation, thus allowing the universe  to transit  from  the  accelerating  phase to  the  radiation-dominated  epoch. In the standard inflationary or cold inflation scenarios,  dissipative effects are typically ignored during the inflationary phase if any pre-inflationary radiation energy density is diluted. The universe then ends up in a supercooling phase requiring a reheating mechanism \cite{Kofman:1994rk}, where the inflaton starts oscillating around the minimum of its potential and progressively dissipates its energy into other relativistic light degrees of freedom, to heat up the universe again as required by the standard big bang cosmology. On the contrary, dissipative effects may be strong enough during inflation where pre-inflationary radiation energy density can be sustained during inflation and also become dominant at the end of inflation whereby the universe smoothly enters into radiation dominated epoch without a  need for a separate reheating period \cite{Berera:1996fm}. Such dissipative effects bring about much richer dynamics for inflation at both background and perturbative levels (for reviews see for eg. \cite{WI Review}) introducing warm inflation as a promising complimentary version of cold inflation by addressing some of long-lasting problems related to (post-)inflationary picture in cold inflation scenarios.

For a comparison with cold inflation it is useful to recall some of the features of warm inflation. It is interesting to note that dissipative effect appears as a supplementary friction term in background equations allowing embedding of steeper potentials in warm inflation solving so-called $\eta$-problem \cite{Berera:1999ws}. Also it leads to several different possibilities for graceful exit depending on the form of potential, form of dissipation coefficient, and whether the dynamics is in strong or weak dissipative regime \cite{Das:2020lut}. Moreover, dissipative effects also modify primordial spectrum of curvature perturbations, resulting in a smaller energy scale of inflation, and reconciling steeper potentials with observational data \cite{Lower-energy scale}. Such  appealing  features  of  warm inflation  allows  it to  simultaneously  satisfy  the  so-called  swampland  conjectures, provided warm inflation can occur in sufficiently strong dissipative regime \cite{WI-swampland, Goswami:2019ehb, Brandenberger:2020oav}. Although it is enormously challenging to achieve a strong dissipative regime in warm inflation, two models were successfully constructed to push warm inflation into strong dissipative regime with inspiration from particle physics \cite{Berghaus:2019whh, Bastero-Gil:2019gao}. Furthermore, the inflaton itself can be a source and responsible for cosmic magnetic field generation \cite{Berera:1998hv} and in combination with the intrinsic dissipative effects lead to a novel dissipative baryogenesis scenario during inflation \cite{WI-baryogenesis}. More recently, it was shown that warm inflation enables a stable remnant of inflaton in the post-inflationary epoch which can behave either like cold dark matter accounting for all the dark matter in the universe \cite{WI-dark matter} or like a quintessence at late time generating the present phase of accelerated expansion \cite{WI-dark energy} (see also \cite{Lima:2019yyv, Sa:2020fvn} for unifying all conventional ingredients of modern cosmology using dissipative effects).

The goal of this manuscript is to investigate the dissipative particle production effects\footnote{For brevity, we label these effects in the following as  dissipative effects. We note that the source of such dissipative effects is particle production.} on pre-inflationary dynamics of $k=1$ LQC and understand their role on the hysteresis-like phenomena and the onset of inflation for Starobinsky potential starting from highly unfavorable initial conditions. Our goal will be to consider those cases which failed to lead to inflation in absence of dissipative effects.  In Sec. II, we  give a brief review of the effective dynamics of  $k=1$ LQC and the warm inflation and discuss the way  dissipative effects are implemented in $k=1$ LQC. In Sec. III, we solve dynamical system of equations in the presence of dissipative effects and show that even small amount of dissipation enlarges the phase space of initial conditions for which inflation occurs. We phenomenologically investigate these solutions for three models of warm inflation:  the warm little inflaton (linear temperature dependent dissipation coefficient), variant of warm little inflaton (inverse temperature dependent dissipation coefficient) and minimal warm inflation (cubic temperature dependent dissipation coefficient). Moreover, we also investigate some features of the qualitative dynamics using phase-space portraits. These results show that in presence of dissipation the hysteresis-like phenomena becomes much stronger and results in a quick onset of inflation for even those initial conditions where inflation could not start in absence of dissipation. We conclude the manuscript with a summary of our results in Sec. IV.

\section{Effective Dynamics in $k=1$ LQC and Warm Inflation}

In this section we first briefly review the effective dynamics of spatially closed loop quantum cosmology in the holonomy quantization \cite{apsv}. This is followed by a discussion of the dynamical equations in the warm inflation scenario and the way warm inflation can be implemented in the effective spacetime description of $k=1$ model in LQC.

\subsection{Effective dynamics of k=1 LQC}
LQC is a canonical quantization based on Ashtekar-Barbero variables -- the connection  $A^i_a$ and its conjugate triad $E^i_a$ which due to homogeneity and isotropy, symmetry reduce to $c$ and $p$ for the $k=1$ Friedmann–Lema\^itre–Robertson–Walker (FLRW) model . In the improved dynamics or the $\bar \mu$ scheme of LQC \cite{aps3}, it turns out that an equivalent set of variables defined as $b = c|p|^{-1/2}$ and $v=|p|^{3/2}$
are more convenient to obtain the quantum and effective description. Here $v$ denotes the physical volume of the unit sphere spatial manifold and is related to the scale factor of the universe as $v = 2\pi^2a^3$. The phase space variables $b$ and $v$ satisfy $\{b,v\}=4\pi G\gamma$ where $\gamma$ denotes the Barbero-Immirzi parameter whose value is generally taken to be $\gamma \approx 0.2375$ in LQC following the calculations of black hole thermodynamics in loop quantum gravity.

The effective Hamiltonian in the holonomy based quantization of the $k=1$ model in LQC for the lapse chosen as unity is given by
\begin{widetext}
\begin{align}
\mathcal{H}_{\mathrm{eff}} = - \frac{3}{8\pi G \gamma^{2} \lambda^{2}} v \left[\sin ^{2}(\lambda b - D) - \sin^{2}D + (1+\gamma^{2})D^{2}\right] + \mathcal{H}_{\mathrm{matt}} \approx 0
\end{align}
\end{widetext}
where $D = \left(\lambda (2\pi^{2})^{1/3}\right))/v^{1/3}$ and $\lambda^{2} = 4(\sqrt{3}\pi\gamma) \ell_{pl}^{2}$. Here we have ignored the modifications from the inverse volume effects which turn out to be negligible in comparison to the holonomy modifications \cite{apsv}.\footnote{Note that a non-singular dynamics results  solely from inverse volume modifications too in $k=1$ model in LQC \cite{topo}, which has been used to understand conditions for the onset of inflation \cite{qm}.} Before we examine the dynamics resulting from this Hamiltonian let us note that there exists another quantization of the $k=1$ model in LQC which is known as the connection based quantization \cite{ck2011}. 
Though there exist some qualitative differences in the way singularity resolution occurs in this prescription as compared to the holonomy based quantization \cite{ck2011, ds2017}, the main features of dynamics remain the same. Especially, the existence of hysteresis which plays an important role in the onset of inflation is robust in both the quantization prescriptions and the difference between the two approaches turn to be small for inflationary dynamics \cite{Dupuy:2019ibu}. For this reason we consider only the effective dynamics for the holonomy quantization in this analysis.

Using Hamilton's equations, the equation of motion for volume turns out to be
\begin{align}
\dot v = \{v, \mathcal{H}_{\mathrm{eff}}\} = \frac{3}{\gamma \lambda} v \sin (\lambda \beta - D) \cos (\lambda \beta - D)
\end{align}
which  results in the following modified Friedmann equation
\begin{align}
H^{2} = \frac{\dot v^{2}}{9v^{2}} = \frac{8\pi G}{3}(\rho - \rho_{\rm{min}})\left(1- \frac{\rho-\rho_{\rm{min}}}{\rho_{\max}^{\rm{flat}}}\right) ~.
\end{align}
Here $\rho_{\max}^{\rm{flat}} = 3/(8\pi G \gamma^{2} \lambda^{2})$ denotes the energy density at the bounce for the spatially-flat model in LQC and
\begin{align}
\rho_{\rm{min}} = \rho_{\max}^{\rm{flat}} \left[(1+\gamma^{2})D^{2}+ \sin ^{2}(D)\right] ~
\end{align}
denotes the minimum allowed energy density in the evolution. In the classical universe this value coincides with the value of energy density at which a classical recollapse occurs. The maximum of the energy density is given by
\begin{align}
\rho_{\rm{max}} = \rho_{\rm{min}} + \rho_{\max}^{\rm{flat}}~.   \label{maxdensity}
\end{align}

Note that in the quantum regime depending on the initial conditions, a bounce as well as a recollapse can occur at $\rho_{\rm{min}}$ as well as at $\rho_{\rm{max}}$ \cite{gls20}.

The Hamilton's equations for  phase space variable conjugate to $v$ is given by
\begin{align}
\dot b = \{b, \mathcal{H}_{\rm{eff}}\} = - 4\pi G \gamma \left[\rho + P - \rho_{1}\right ]
\end{align}
with
\begin{align}
\rho_{1} = \frac{\rho_{\max}^{\rm{flat}}D}{3} \left[2(1+\gamma^{2})D - \sin (2\lambda \beta - D) - \sin (2D)\right]
\end{align}
where $P$ denotes the pressure which equals $P = - \partial \mathcal{H}_{\mathrm{matt}}/{\partial v}$. The dynamical equations for the scalar field matter variables with a potential $V(\phi)$ are
\begin{align}
\dot \phi &= \{\phi, \mathcal{H}_{\rm{eff}}\} = \frac{p_{\phi}}{p^{3/2}}\\
\dot p_{\phi} &= \{p_{\phi}, \mathcal{H}_{\rm{eff}}\}= - p^{3/2} V_{,\phi} ~.
\end{align}
Using above equations it is straightforward to show that Klein-Gordon equation follows along with the standard conservation law for matter energy density.

Above dynamical equations encode non-perturbative quantum gravitational effects which result in a non-singular bounce of the universe in the Planck regime \cite{apsv,Dupuy:2019ibu,gls20}. This results in non-singular cycles of expansion and contraction if the matter does not violate strong energy condition i.e. has equation of state $\omega = P/\rho$ greater than $-1/3$. For latter type of matter content, the universe undergoes a  recollapse at late times resulting in a contraction and a big crunch singularity in the classical theory. This singularity is avoided in LQC resulting in a bounce and another phase of expansion and a possible recollapse if the equation of state $w > -1/3$ in the expanding phase.
If the recollapses occur at the macroscopic scales,
the difference in the volumes of two consecutive recollapses is found to be \cite{Dupuy:2019ibu}
\bq
\lb{changeinvolume}
\delta v^{1/3}_\mathrm{rec}=\frac{-\oint P dv}{(2\pi^2)^{2/3}\rho_\mathrm{max}^{\rm{flat}}\gamma^2\lambda^2}.
\eq
This implies that in each cycle of expansion and contraction the maximum volume $v_{\mathrm{rec}}$ changes. This occurs because of the asymmetry of the pressure during different phases of a given cycle which results in a
 hysteresis-like phenomena  \cite{Sahni:2012er,Dupuy:2019ibu}. This hysteresis-like phenomena has been shown to be responsible for alleviating problems with onset of inflation for different potentials, especially low energy scale models \cite{gls20}. Before we examine this phenomena in presence of radiation production in warm inflationary scenarios, we summarize the latter and obtain the relevant equations in LQC.

\subsection{Warm inflationary dynamics in LQC}

The dynamical realization of warm inflation is different from the cold inflation due to the presence of radiation as well as the possibility of energy exchange between inflaton and radiation energy density. Hence, the total energy density of the universe in warm inflation reads
\begin{align}
\rho = \rho_{\phi} + \rho_{r}
\end{align}
where $\rho_{\phi} = \frac{1}{2} \dot \phi^{2} + V(\phi)$ is scalar field energy density with $V(\phi)$ being some potential function and $\rho_{r}$ is the radiation energy density. The inflaton field $\phi$ and the radiation energy density form a coupled system in warm inflationary dynamics due to dissipation of energy out of the inflaton system and into radiation. The background evolution equations are respectively given  by \cite{Das:2020lut}
\begin{align}
&\ddot \phi + 3H\dot \phi + V_{,\phi} = -\Upsilon (\phi, T) \dot \phi\\
&\dot\rho_{r} + 4H \rho_{r} = \Upsilon (\phi, T) \dot\phi^2 . \label{radiation}
\end{align}
Here $\Upsilon(\phi, T)$ is the dissipation coefficient which can be a function of both inflaton and temperature, depending  on  the  specifics  of  the microscopic physics behind the construction of a warm inflation model.  For a
radiation or a bath of relativistic particles, the radiation energy density is given by $\rho_{r} = \left(\pi^2 g/30\right) T^4$, where $g$ is the effective number of light degrees of freedom ($g$ is fixed according to the dissipation regime and interactions form used in warm inflation). Such radiation production results in entropy production where the entropy density $s$ is related to radiation energy density by $Ts = (4/3) \rho_{r}$, i.e., is related to temperature as $s = \left(2 \pi^{2} g/45\right) T^3$, where we have considered a thermalized radiation bath as is typically the case in warm inflationary scenarios. Then, eq. (\ref{radiation}) can be written in terms of entropy as follows \cite{Brandenberger:2020oav}
\begin{align}
T\left(\dot s + 3Hs\right) = \Upsilon \dot \phi^{2}.
\end{align}

As we will see in next section, such entropy production significantly changes the hysteresis-like phenomena. In fact, the term $3Hs$, which is positive in expanding  universe ($H>0$) and negative in contracting universe ($H<0$), produces a larger difference in pressure during expansion/contraction stages making hysteresis-like phase stronger in comparison with the case without dissipative effects.

Let us note that the richer dynamics of warm inflation sharpened the interest for finding explicit models aiming at overcoming two important issues found in earlier particle physics realizations of warm inflation.  First, the requirement of large field multiplicities so as to be able to sustain a nearly-thermal bath and, second, the difficulty to achieve strong dissipative regimes ($\Upsilon \gg H$), due to the interplay between inflaton and radiation fluctuations, leading to appearance of growing modes in the scalar curvature power spectrum and that can render it inconsistent with the observations. The former problem was first solved with an introduction of a new class of warm inflation model building realization   motivated from the ingredients used in ``Little Higgs'' models of electroweak symmetry breaking where  the  inflaton  is  a  pseudo-Nambu  Goldstone  boson  of  a  broken  gauge  symmetry and its potential is protected against large radiative corrections by symmetry obeyed by the model while still having enough interactions to allow thermalization of light degrees of freedom. This results in enough dissipation even if the mediators are very light with respect to ambient temperature. In such a model also known as warm little inflaton, the dissipation coefficient is given by \cite{Bastero-Gil:2016qru}
\begin{align}
\Upsilon_{\mathrm{lin}}  = C_{\mathrm{lin}} T.
\end{align}
We  refer  to  the above  $\Upsilon_{\mathrm{lin}}$ as  the \textit{linear  dissipation  coefficient}. Although warm little inflaton was successful in producing sustainable thermalized radiation bath utilizing just a few mediator fields, it could not obtain strong dissipative regime, which allows steeper potentials to be embedded in warm inflation by making energy scale of inflation smaller. To this end, a concrete model of warm inflation, the so  called minimal warm inflation \cite{Berghaus:2019whh}, was recently constructed in which the inflaton has axion-like coupling to gauge fields. Since the inflaton is an axion, its shift symmetry protects it from any perturbative backreactions  and  thus  from  acquiring  a  large  thermal  mass. Hence, the thermal friction from this bath can easily be stronger than Hubble friction even for small number of fields. The corresponding axion friction coefficient turn out to be as
 \begin{align}
\Upsilon_{\mathrm{cub}}  = C_{\mathrm{cub}} T^{3}.
\end{align}
Hereafter,  we  refer  to  the above  $\Upsilon_{\mathrm{cub}}$ as  the \textit{cubic  dissipation  coefficient}. In this regard, another model was also recently proposed inspired from an idea used in warm little inflaton where the inflaton is  directly coupled to  light  scalar  bosonic  fields rather than fermionic fields which is known as a variant of warm little inflaton \cite{Bastero-Gil:2019gao}. Although the exact form of dissipation coefficient is complex, the leading behaviour of dissipation, when the effective mass is dominated by thermal part, varies as
\begin{align}
\Upsilon_{\mathrm{inv}}  = C_{\mathrm{inv}} T^{-1}.
\end{align}
Hereafter,  we  refer  to  the above  $\Upsilon_{\mathrm{inv}}$ as  the \textit{inverse  dissipation  coefficient}. We should note that $C_{\mathrm{inv}} \ll C_{\mathrm{lin}} \ll C_{\mathrm{cub}}$ since it should be fixed in such a way that the condition for sustainable thermal bath, i.e. $T>H$, is satisfied during inflationary phase.

Taken together, to consider the dissipative effects during both pre-inflationary and inflationary phases all the way from the bounce until the end of inflation, we phenomenologically implement dissipative effects into the effective equations of spatially closed model LQC. The resulting dynamical equations are, \begin{align} \label{dynamical equations}
\dot v & = \frac{3}{\gamma \lambda} v \sin (\lambda b - D) \cos (\lambda b - D)\\
\dot b & = - 4\pi G \gamma \left[ \frac{p_{\phi}^{2}}{v^{2}} + \frac{4}{3} \rho_{r} - \rho_{1}\right]\\
\dot \phi & = - \frac{p_{\phi}}{v}\\
\dot p_{\phi} & = - v \, V_{,\phi} - \Upsilon(\phi, T) \, p_{\phi}\\
\dot \rho_{r} & = -\left[\frac{4}{\gamma \lambda} \sin (\lambda b - D) \cos (\lambda b - D)\right] \rho_{r} + \frac{\Upsilon(\phi,T) \, p_{\phi}^{2}}{v^{2}}.
\end{align}

In the next section, we will first discuss the way such dissipative effects, or equivalently entropy production, change the dynamics of pre-inflationary phase and also enlarge the phase-space of initial conditions which result in a (warm) inflation. Then, we perform a qualitative analysis of the dynamical equations to understand the attractor behavior of the solutions and gain insights on the way dissipative effects help in the onset of inflation even starting from highly unfavorable initial conditions.

\section{Dissipative effects on pre-inflationary dynamics of k=1 LQC}

In this section we investigate the consequences of dissipative effects in LQC to address the problem of onset of inflation for the Starobinsky potential.  We discussed in Sec. 2 the way non-singular cycles of expansion and contraction result in a hysteresis-like phenomena which arises due  to
differences in pressure during expansion and contraction stages of cosmic evolution. Due to this difference in pressure, the work done during one cycle can be positive or negative depending on the potential function. For sufficiently flat potentials, the work can be positive resulting in increasing the size of the universe in the successive cycles. Because of this even if the inflaton starts with a kinetic energy dominated conditions and an equation of state close to unity, the equation of state decreases in each cycle and eventually becomes less than $-1/3$ which leads to an onset of inflation.   Hence, if the universe fails to inflate after the first cycle, it can do so after subsequent cycles enlarging the phase-space of initial conditions which results in inflation \cite{Dupuy:2019ibu}. Recently it was shown that for Starobinsky potential universe can inflate for a large part of initial conditions, however, it should go though numerous cycles of expansion and contraction \cite{gls20}. Further, for some of the initial conditions the inflation does not commence even after a large number of non-singular cycles. As we will see
the dissipation or entropy production, leads to larger differences in pressure during expansion-contraction phase resulting in larger amplitude of the cycles. Therefore, we expect that entropy production due to radiation particle production make the hysteresis phenomena stronger leading into the universe with bigger size in successive cycles, causing the universe to inflate  after small number of cycles. In the following we first obtain the background solutions demonstrating above phenomena which is followed by discussion of phase space portraits in qualitative dynamics of this model.

\subsection{Dissipative effects for Starobinsky potential}
Starobinsky inflation is a prominent example of low energy inflation models favored by current observations. In classical cosmology, this model results from adding $R^2$ term to action which translates to adding the following potential in the Einstein frame
\begin{align}
U(\phi) = \frac{3m^{2}}{32\pi}\left(1- e^{- \sqrt{\frac{16\pi}{3}}\phi(t)}\right)^{2} ~.
\end{align}
But in LQC, the above potential  is not obtained from an $R^2$ term in the action, since the covariant action in LQC does results in higher order curvature terms but in  a Palatini framework \cite{olmo-ps}. As in previous works in LQC, we consider above potential as a phenomenological input in effective dynamics.

In Starobinsky model the inflation is supposed to start at energy scales far lower than Planck scale and as a result the initial conditions in the Planck regime are such that kinetic energy dominates the potential energy. If one numerically solves the classical cosmological dynamics of above potential one finds that the universe undergoes a recollapse before potential energy can dominate and encounters a big crunch singularity \cite{lindelecture, linde2018}. We would see that this situation changes dramatically in the effective dynamics in LQC. Below we   numerically solve dynamical equations (\ref{dynamical equations}) for various initial conditions,  using explicit Runge-Kutta algorithm and stiff-switching method in Mathematica with accuracy and precision goals set to eleven. The initial value of $b$ (the conjugate to volume $v$) is fixed by the vanishing of the effective Hamiltonian constraint. Moreover, we also set the initial value of $p_{\phi}$ using the condition for the bounce, i.e. $\rho = \rho_{\mathrm{max}}$. Therefore, we are left with just three initial conditions on volume ($v_0$), scalar field ($\phi_0$) and initial radiation energy density ($\rho_{r0}$). We choose initial conditions such that the radiation energy density is sub-dominated in comparison with both the kinetic energy density and potential energy density of the inflaton field and the bounce happens with kinetic dominated initial conditions ($\rho_{r0}\ll U(\phi_{0})\ll \dot\phi_{0}^{2}/2$).

\begin{figure}[tbh!]
\includegraphics[scale=0.33]{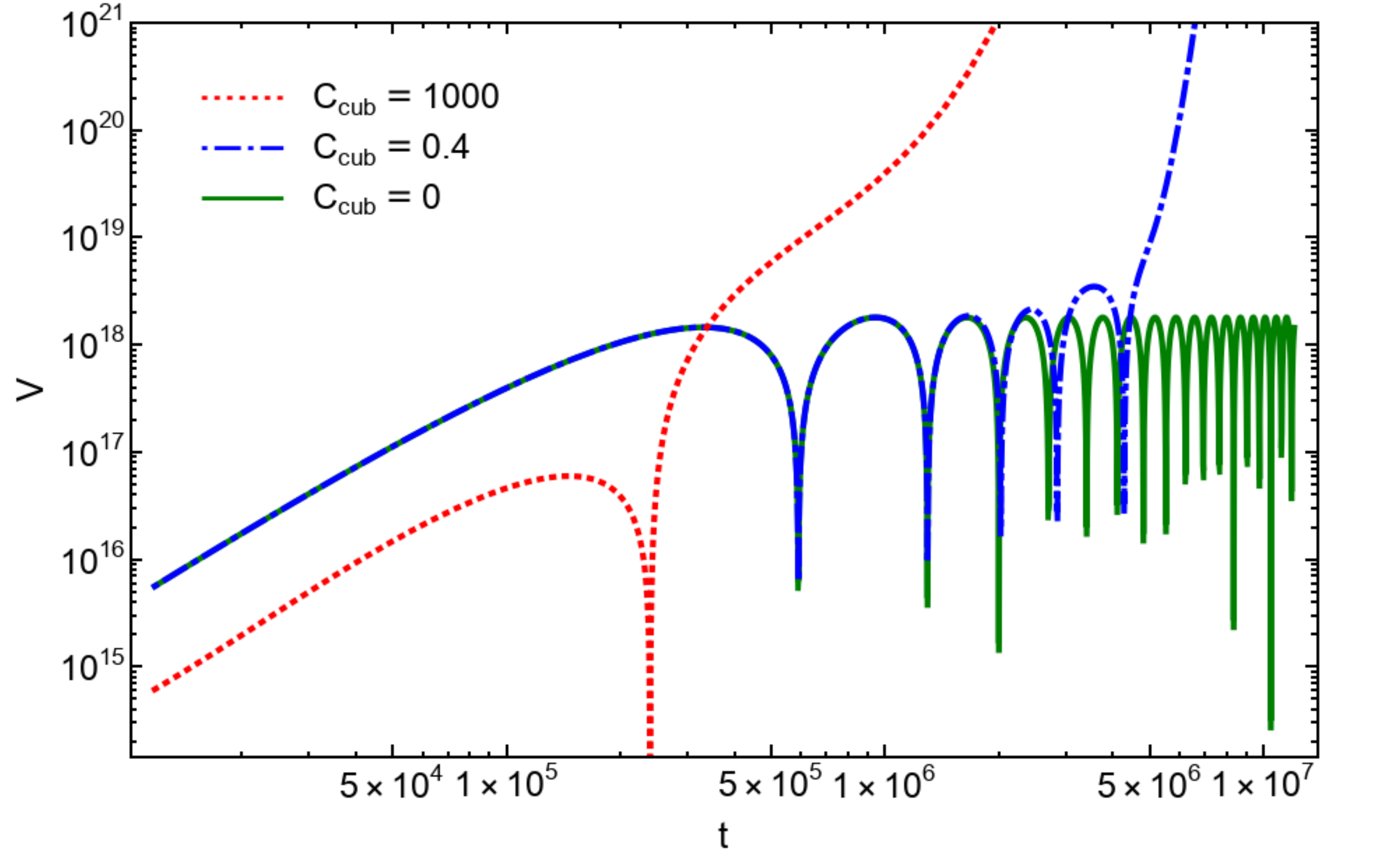}
\caption{The evolution of volume for different values of cubic dissipation coefficient. Initial conditions are chosen at the bounce with $v_{0} = 5 \times 10^7$, $\phi_{0} = -1$, $\rho_{r0} = 10^{-12}$, and $g = 17$.}\label{Vcubic}
\end{figure}

\begin{figure}[tbh!]
\includegraphics[scale=0.33]{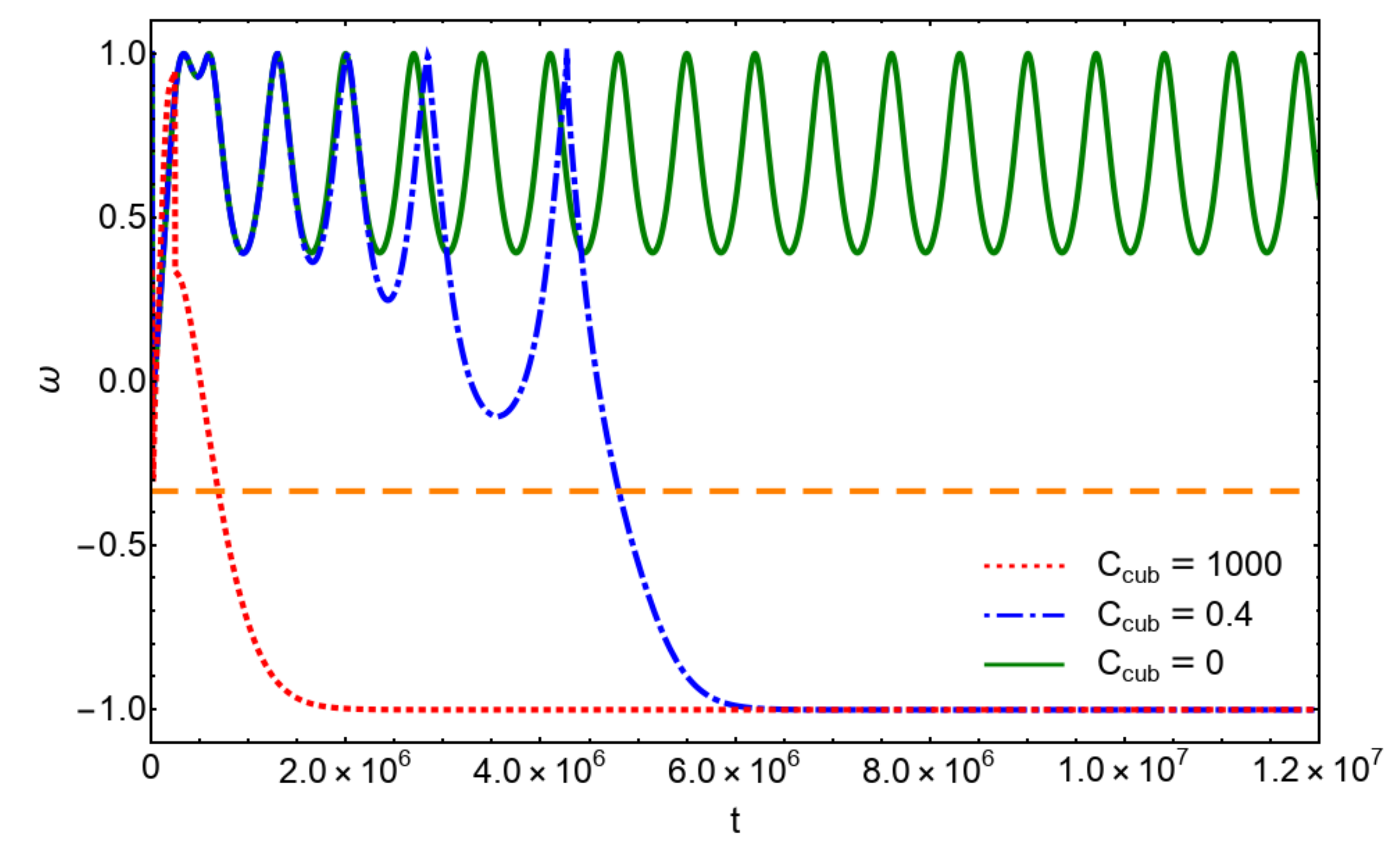}
\caption{The evolution of equation of state for cubic dissipation coefficient and same initial conditions as in Fig. \ref{Vcubic}.} 
\label{Wcubic}
\end{figure}

In the following we first solve the dynamical equations for the case of minimal warm inflationary model i.e. with a cubic dissipation coefficient. In this case  inflaton has an axionic coupling to a non-abelian gauge theory and the sphaleron  transitions  between  gauge  vacua,  existing  at sufficiently high temperatures. And,  if the corresponding non-Abelian gauge  theory  has  gauge  group $SU(3)$,  there  are  8 gauge bosons, each of which  contribute two relativistic degrees of freedom. Including the inflaton itself, there are in total 17 relativistic degrees of freedom. So we set the number of relativistic degree of freedom $g = 17$ \cite{Goswami:2019ehb}. In Figs. \ref{Vcubic} and \ref{Wcubic}, we plot the evolution of volume and equation of state for three different values of $C_{\mathrm{cub}}$ in LQC. For the initial conditions $v_{0} = 5 \times 10^7$, $\phi_{0} = -1$, $\rho_{r0} = 10^{-12}$ the dynamical evolution is non-singular for all the considered values of $C_{\mathrm{cub}}$. The initial conditions are chosen such that in absence of dissipation, inflation does not start after various cycles of non-singular evolution. As can be seen, in case of $C_{\mathrm{cub}}=0$ (dissipationless universe), the universe does not enter an inflationary phase even after many cycles of non-singular evolution.  This is because the hysteresis-like phenomena is not large enough to set scalar field at the flat part of potential function in subsequent cycles. However, dissipative effects make hysteresis phase stronger (decreasing the number of cycles and increasing its amplitudes) whereby the universe begins inflationary phase after small number of cycles. This is evident in the dynamical evolution for $C_{{\mathrm{cub}}} = 0.4$ and $C_{\mathrm{cub}} = 1000$. We see from the former case that even a small non-zero value of $C_{\mathrm{cub}}$, resulting in small dissipative effects, has substantial effects on the hysteresis-like phenomena and the onset of inflation. We find from the volume and equation of state plot that a phase of inflation starts after a few non-singular cycles when volume grows exponentially and equation of state becomes less than $-1/3$. But such a small value of dissipation coefficient can not sustain the thermal bath during inflation and one needs a larger value of $C_{\mathrm{cub}}$. As one increases the value of $C_{\mathrm{cub}}$, the dissipative effects make hysteresis phase very strong and inflationary phase starts just after just one bounce. This is shown in Fig. \ref{Vcubic} for the case of $C_{\mathrm{cub}} = 1000$. Here we should note that the curves for $C_{\mathrm{cub}} = 0.4$ and $C_{\mathrm{cub}} = 1000$ start from the same initial volume but because of the use of  logarithmic scale in the plot, the figure does not show the same value of volume for both the curves. We note that the evolution of equation of state in Fig. \ref{Wcubic} shows that for $C_{\mathrm{cub}} = 0$, the equation of state oscillates between 1 and 0.5 for the entire range of evolution, however for non-vanishing dissipation coefficients it decreases quickly below $w = -1/3$ and becomes $w \approx -1$ indicating an onset of slow-roll inflation. As one can see, the equation of state becomes $-1$ much earlier for larger dissipation coefficient.

\begin{figure}
\includegraphics[scale=0.33]{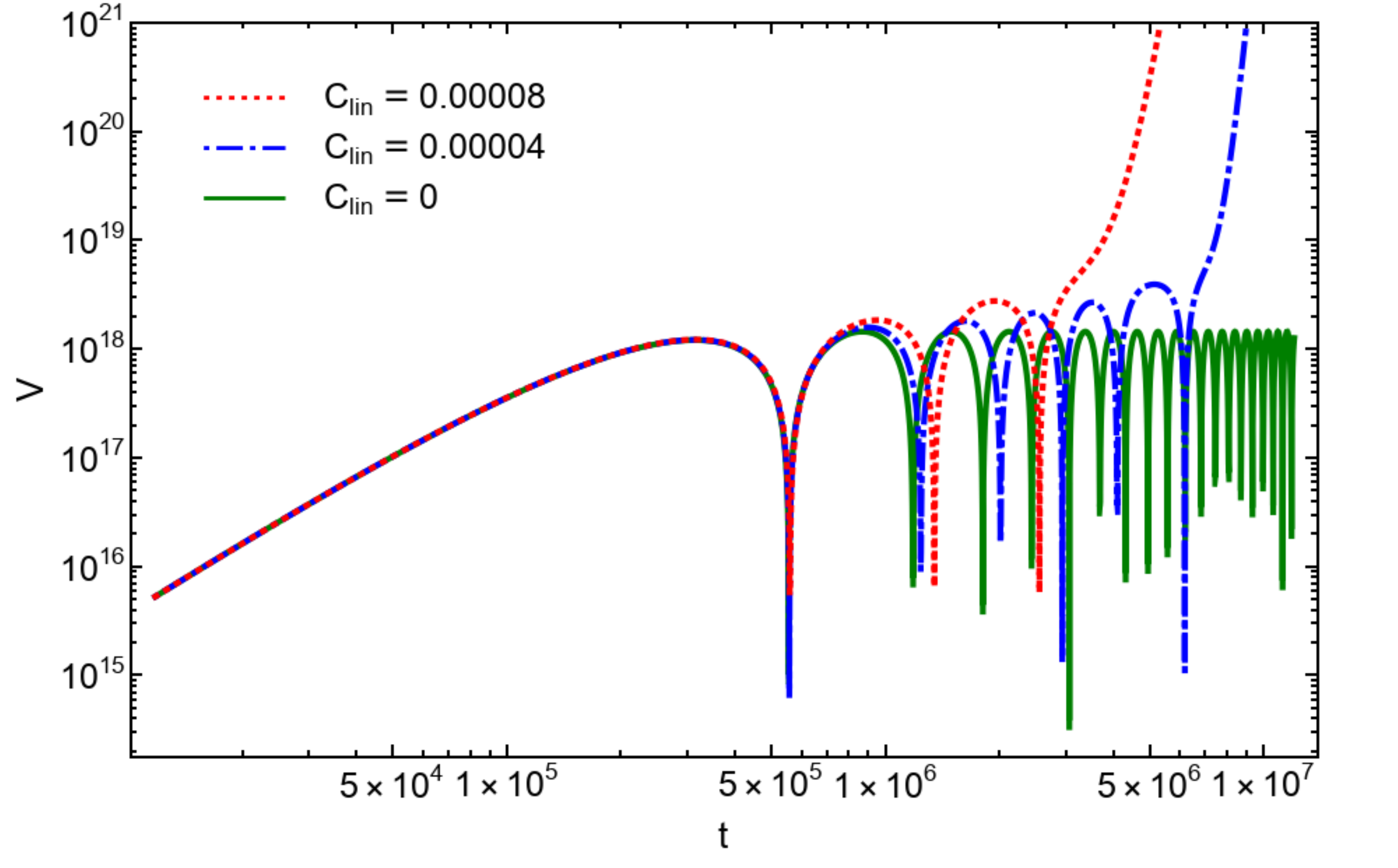}
\caption{The evolution of volume for different values of linear dissipation coefficient. Initial conditions are chosen at the bounce with $v_{0} = 10^7$, $\phi_{0} = -1.5$, $\rho_{r0} = 10^{-11}$, and $g = 12.5$.}\label{Vlinear}
\end{figure}

\begin{figure}
\includegraphics[scale=0.33]{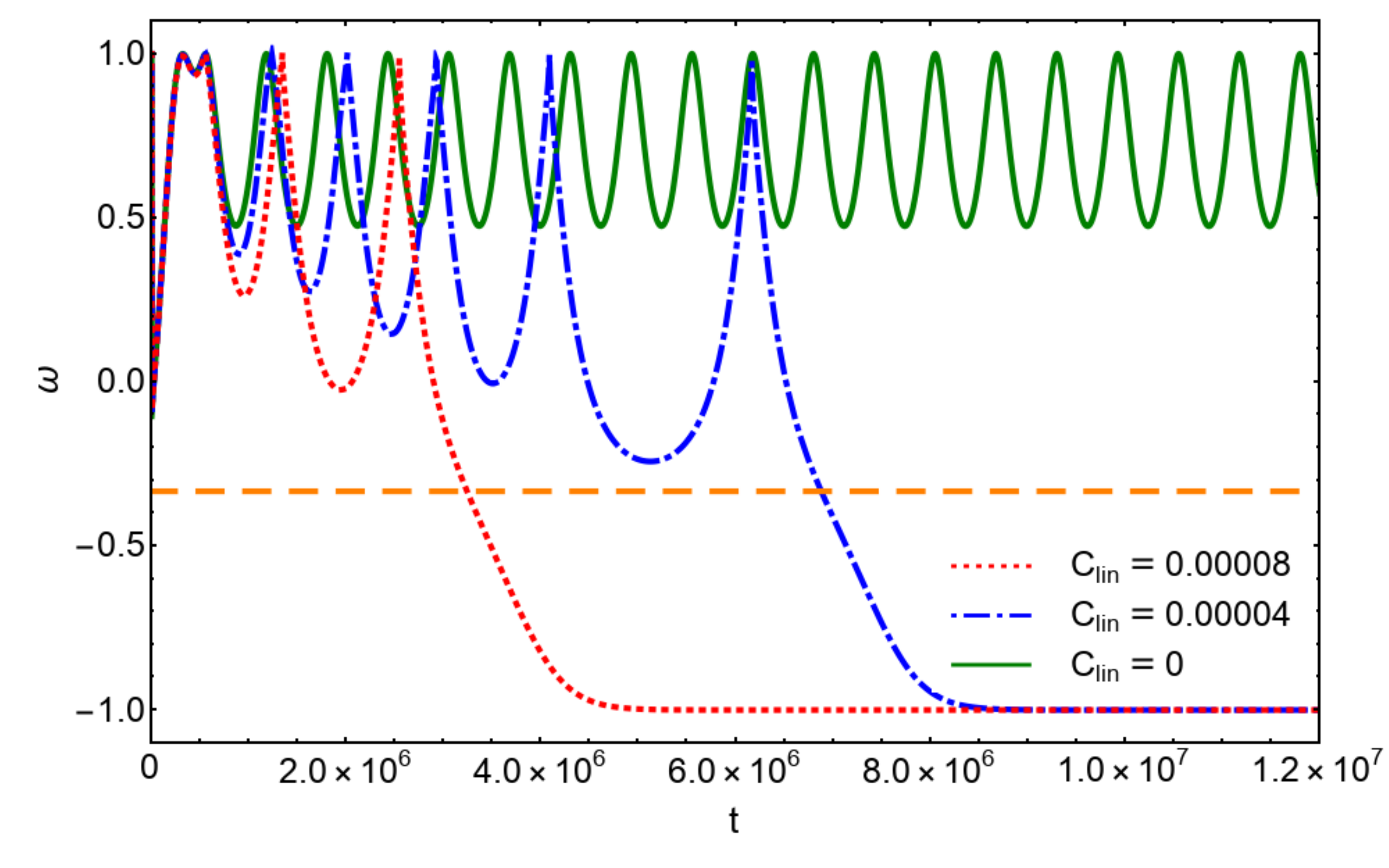}
\caption{The evolution of equation of state for linear dissipation coefficient and same initial conditions as in Fig. \ref{Vlinear}.}
\label{Wlinear}
\end{figure}

We now discuss the case of warm little inflaton in LQC.
In Figs. \ref{Vlinear} and \ref{Wlinear}, we plot evolution of volume and equation of state for linear dissipation coefficient and three different values of $C_{\mathrm{lin}}$ for initial conditions $v_{0} = 10^7$, $\phi_{0} = -1.5$, and $\rho_{r0} = 10^{-11}$. We also fix $g=12.5$ using analysis in \cite{Bastero-Gil:2016qru}. We consider non-zero values of $C_{\mathrm{lin}}$ as 0.00004 and 0.00008 which are typical values for warm inflation to happen in spatially-flat spacetime.
As before, the chosen initial conditions correspond to the unfavorable ones where inflation does not start in LQC even after various cycles of non-singular evolution when dissipation is absent. This can be seen from the curve corresponding to $C_{\mathrm{lin}}=0$, where universe oscillates in non-singular evolution but there is no onset of inflation since equation of state never becomes less than $-1/3$. However, when we add dissipative effects, the hysteresis becomes stronger and we see that the universe experiences an inflationary phase after a small number of cycles. We see that the equation of state becomes less than $-1/3$ after a few cycles for $C_{\mathrm{lin}}=0.00004$ and $C_{\mathrm{lin}}=0.00008$. As we increase the value of $C_{\mathrm{lin}}$, the number of cycles prior to onset of inflation decrease and the amplitude of the cycles become larger.
\begin{figure}
\includegraphics[scale=0.33]{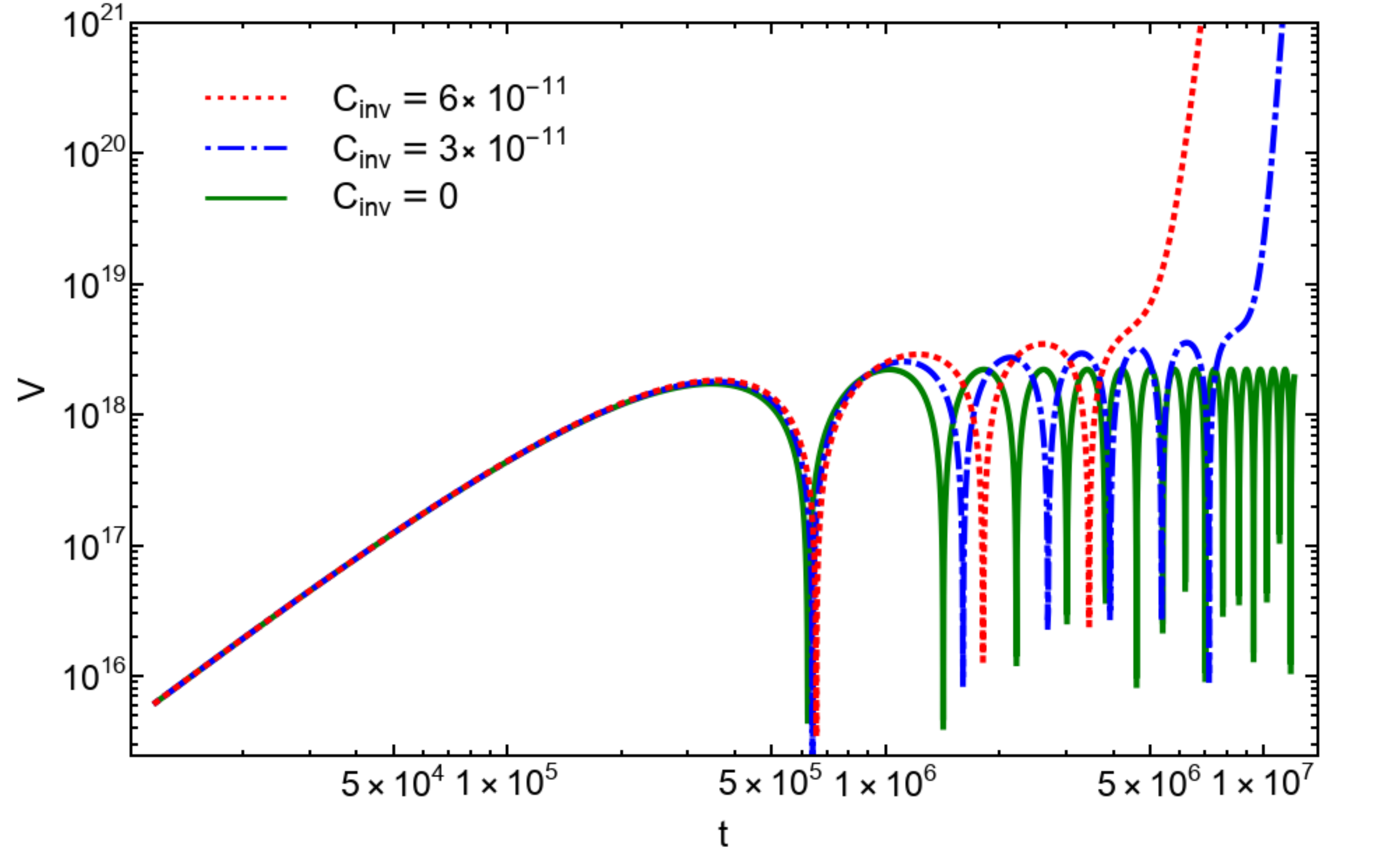}
\caption{The evolution of volume for different values of inverse dissipation coefficient. Initial conditions are chosen at the bounce with $v_{0} = 2.5 \times 10^6$, $\phi_{0} = -2$, $\rho_{r0} = 10^{-9}$, and $g = 12.5$.}\label{Vinverse}
\end{figure}

\begin{figure}
\includegraphics[scale=0.33]{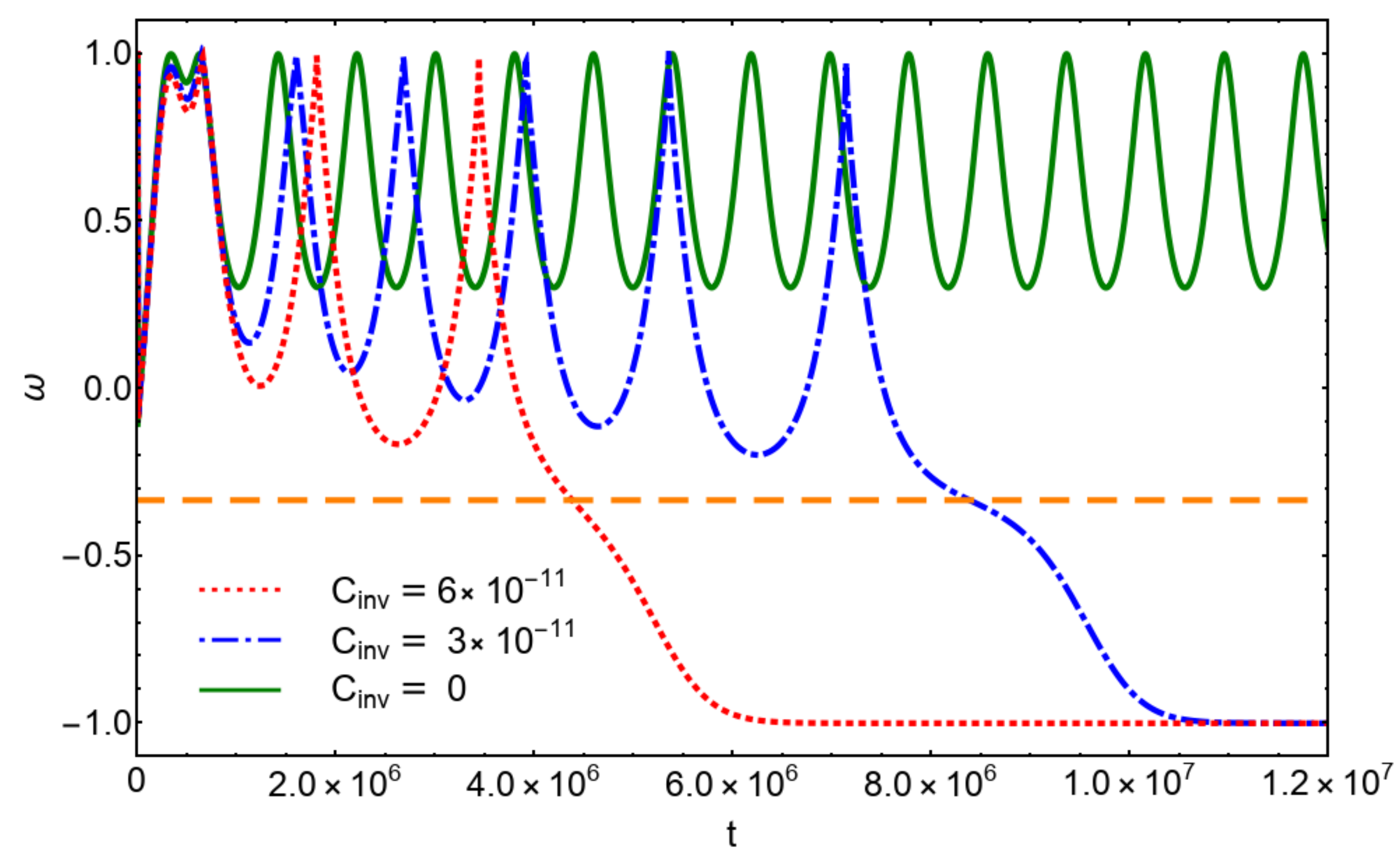}
\caption{The evolution of equation of state for inverse dissipation coefficient and same initial conditions as in Fig. \ref{Vinverse}.}\label{Winverse}
\end{figure}

Finally, we consider the variant of warm little inflaton with inverse dissipation coefficient which is shown in  Figs. \ref{Vinverse} and \ref{Winverse}. As in previous cases, initial conditions are chosen such that there is no inflationary phase even after many cycles in absence of dissipative effects. But choosing a non-zero dissipation coefficient, even if small, leads to a striking difference in dynamics and results in a stronger phenomena of hysteresis. In these figures we choose  $C_{\mathrm{inv}} = 3 \times 10^{-11}$ and $6 \times 10^{-11}$ and fix  $g=12.5$. The chosen values of $C_{\mathrm{inv}}$ are smaller than two other cases as we discussed previously.  We find that as we increase the value of $C_{\mathrm{inv}}$, the number of cycles decrease and the amplitude of the cycles become larger and the universe enter into inflationary phase sooner.

To summarize the results so far,
we have found the dissipative effects resulting in radiation production, make the hysteresis phenomena stronger and setting the condition for inflation to happen sooner for dissipation coefficients which have cubic, linear and inverse relationship to temperature. Though we discussed a sample of initial conditions, our results are robust to changes in initial conditions. To gain some insights on the qualitative dynamics and robustness of results we study the phase-space portraits in the following.

\subsection{Qualitative dynamics in phase-space portrait}
It is useful to understand the phase space portraits for qualitative dynamics by introducing following variables,
\begin{align}
X(t) &= \chi_{0} \left(1- e^{-\sqrt{\frac{16\pi G}{3}}\phi(t)}\right)\\
Y(t) &= \frac{p_{\phi}(t)}{v(t)\sqrt{2\rho_{\mathrm{max}}}} ~~
Z(t) = \sqrt{\frac{\rho_{r}(t)}{\rho_{\mathrm{max}}}}
\end{align}
where $\chi_{0} = m \sqrt{\frac{3}{32\pi G \rho_{\mathrm{max}}}}$
and $\rho_{\mathrm{max}}$ denotes the maximum energy density (\ref{maxdensity}) determined by the initial conditions.  Our goal will be to find the inflationary attractors for different choices of dissipation coefficients and initial conditions. These have been studied earlier for cold inflation in detail in LQC \cite{Li:2018fco}. The inflationary attractor lies at $(X = 0, Y = 0)$ which corresponds to the reheating phase in cold inflation and the beginning of radiation epoch in warm inflation.

\begin{figure}
\includegraphics[scale=0.45]{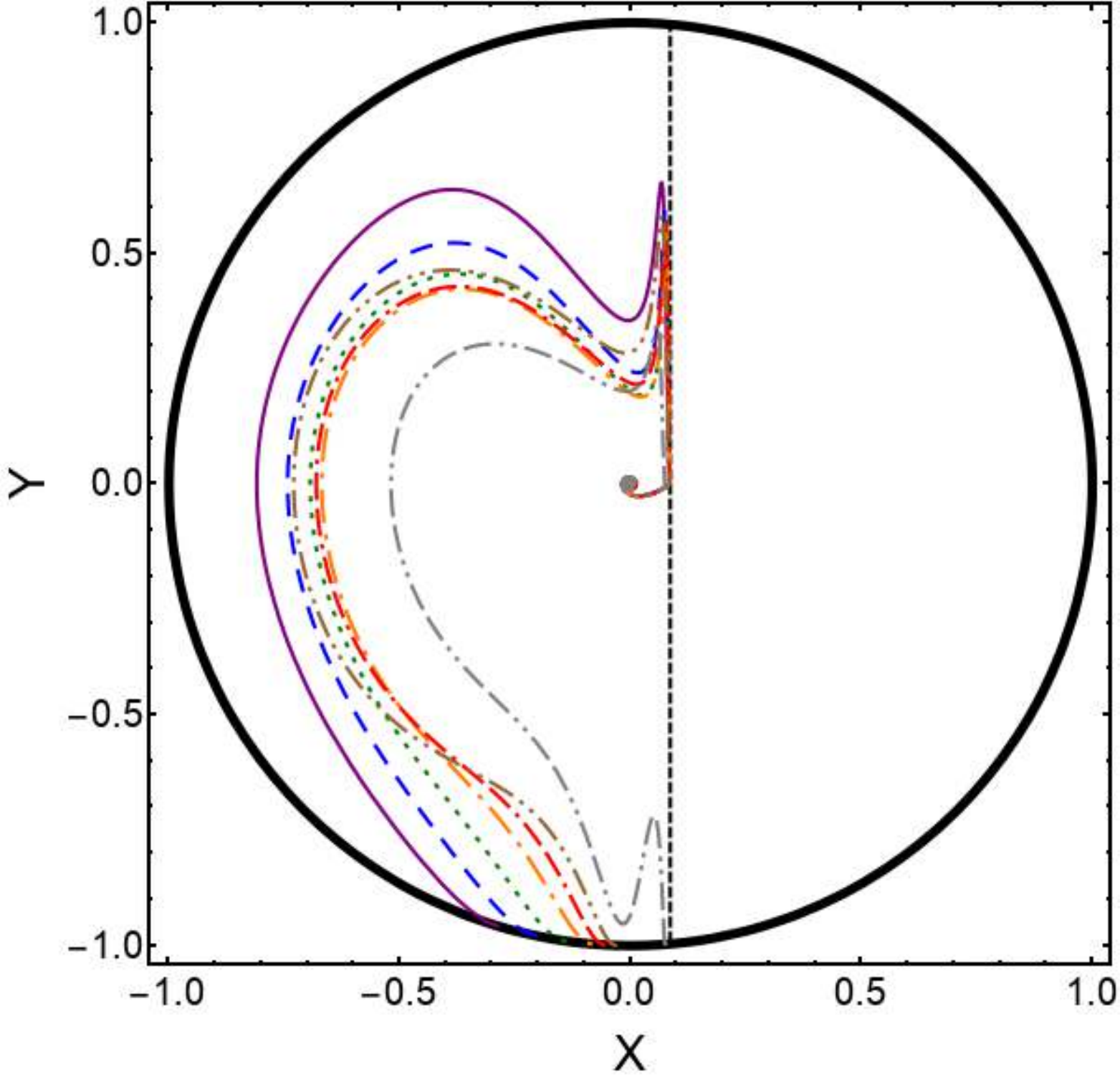} \ \ \ \ \ \ \ \ \ \
\includegraphics[scale=0.47]{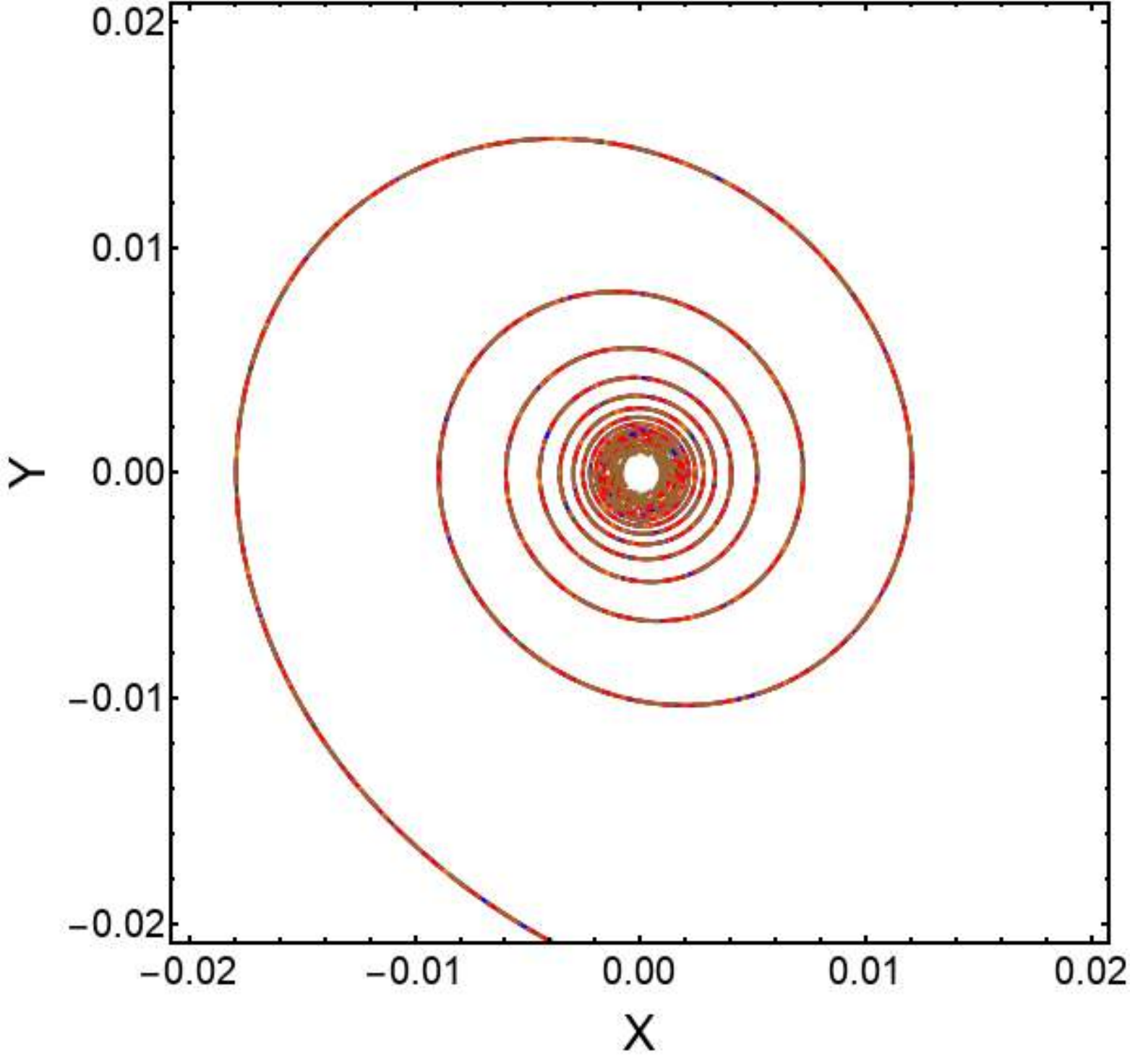}
\caption{Projection of 3 dimensional phase space portrait on $Z=0$ plane for cubic dissipation coefficient with $m=0.62$, $C_{\mathrm{cub}} = 7.5$, $v_{0} = 35$, $\rho_{r0}=10^{-3}$ and seven distinct initial conditions for $\phi_{0}$.}
\label{Phase-Space-cubic}
\end{figure}
\begin{figure}
\includegraphics[scale=0.3]{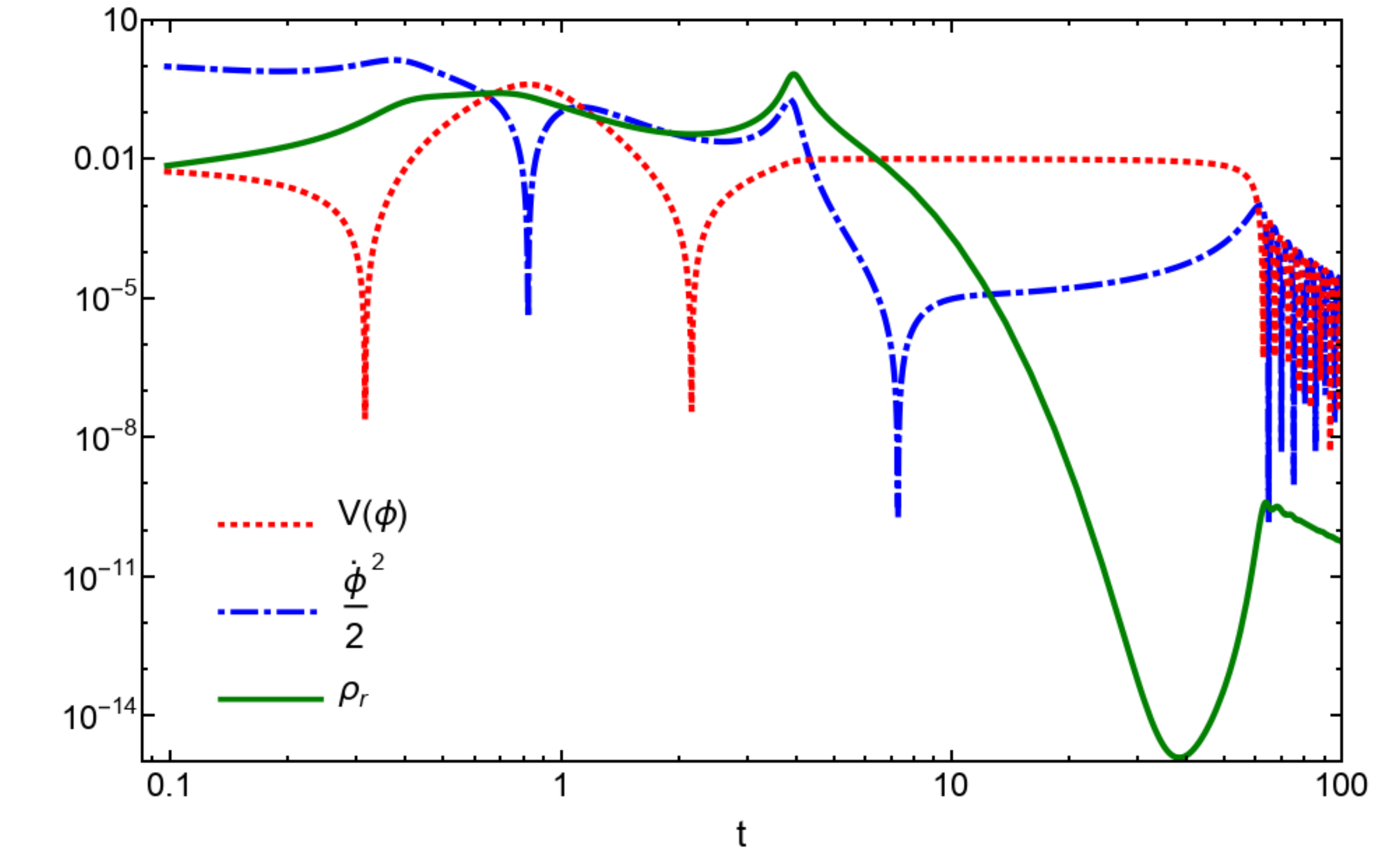}
\includegraphics[scale=0.36]{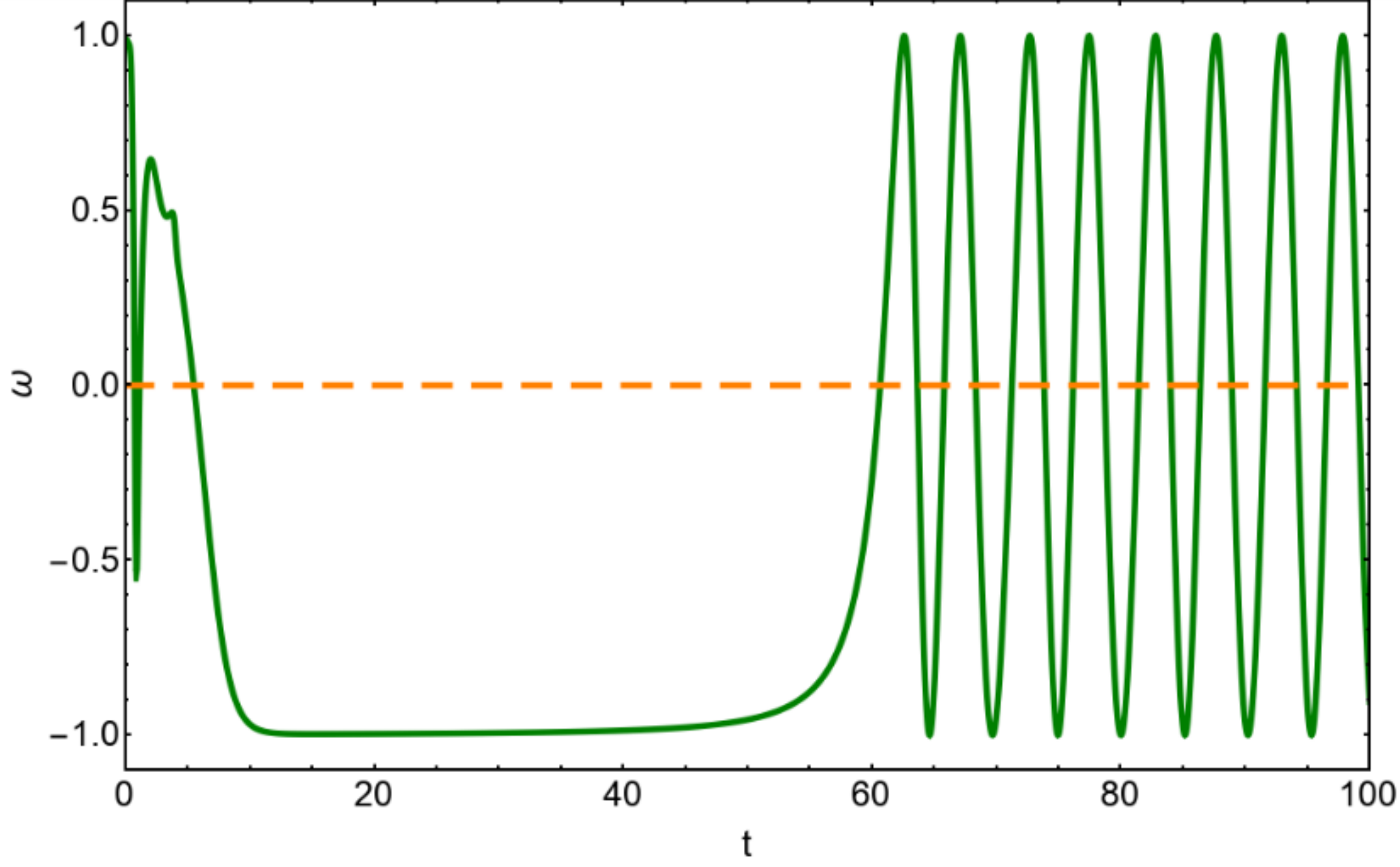}
\caption{Evolution of energy components as well as equation of state for cubic dissipation coefficient with $m=0.62$, $C_{\mathrm{cub}} = 7.5$, $v_{0} = 35$, $\rho_{r0}=10^{-3}$ and $\phi_{0} = 0.5$ (gray curve in Fig. \ref{Phase-Space-cubic}).}
\label{EWcubic}
\end{figure}

Starting with the cubic dissipation in Fig. \ref{Phase-Space-cubic}, the left plot shows the projection of entire phase space region on plane $Z=0$ and the right plot zooms in on the attractor near the origin. The vertical dashed black line corresponds to $X = \chi_{0}$ and all real solutions lie to the left of this line. For a better visualization of the qualitative features it is useful to consider a large value of inflaton mass which is here chosen to be  $m=0.62$. The solid black circular curve corresponds to the energy density of the first bounce at $t = 0$ where the initial conditions are set. The left plot in Fig. \ref{Phase-Space-cubic} shows curves corresponding to seven distinct initial conditions for $\phi_{0}$, while other parameters are fixed, for which the universe undergoes inflation in presence of dissipative effects. For the same initial conditions in absence of dissipative effects the universe goes through many cycles without onset of an inflationary phase.   It should be noted that we chose $C_{\mathrm{cub}}$ large enough to see the end of inflation and also post-inflationary phase. So in most of the cases, the hysteresis-like phenomena go away and universe experience an inflationary phase after just one bounce.

We expand on details of above dynamics in  Fig. \ref{EWcubic}, where the evolution of the potential energy density $V(\phi)$, kinetic energy density $\dot\phi^{2}/2$ and the radiation energy density $\rho_{r}$ are shown for gray curve in Fig. \ref{Phase-Space-cubic}. We can see from the figure that the universe starts from a bounce in the kinetic dominated regime (initial condition with equation of state $\omega \approx 1$), with $\dot\phi^{2}/2\gg U(\phi)\gg \rho_{r}$, and the kinetic energy very soon dilutes away since it behaves as $a^{-6}$. In the subsequent evolution we find that radiation energy density becomes important in comparison to kinetic and potential energies. Such radiation dominated regime before the inflationary phase has also been reported for warm inflation in spatially-flat LQC \cite{Graef:2018ulg} (see also \cite{warm-LQC} for review on warm inflation in spatially-flat LQC). After some cycles, when the hysteresis-like phenomena causes an onset of inflation, the radiation energy density becomes subdominant in evolution.
  Note that contrary to cold inflation, radiation is concurrently produced during inflationary phase and it may reach an equality with potential energy density at the end of inflation if dissipative effects are strong enough to sustain the thermal bath, and the universe smoothly enters into radiation dominated epoch without subsequent (p)reheating phase. However, as it is clear in Fig. \ref{EWcubic}, the universe does not enters into radiation dominated epoch at the end of inflation. This is because the considered value of $C_{\mathrm{cub}}$, chosen due to computational constraints,  is not large enough and hence the inflationary phase is a cold-type and a reheating mechanism is a must.
  Hence, although small dissipation coefficient could not sustain thermal bath during inflation leading to warm-type inflation but it has substantial effect on pre-inflationary dynamics and the onset of inflation which is evident from the existence of a $\omega \approx -1$ phase from the plot of equation of state. We further note that if dissipation coefficient is chosen large enough the warm inflation starts quickly and the radiation energy density reaches similar values as the potential energy density at the end of inflation.

\begin{figure}
\includegraphics[scale=0.45]{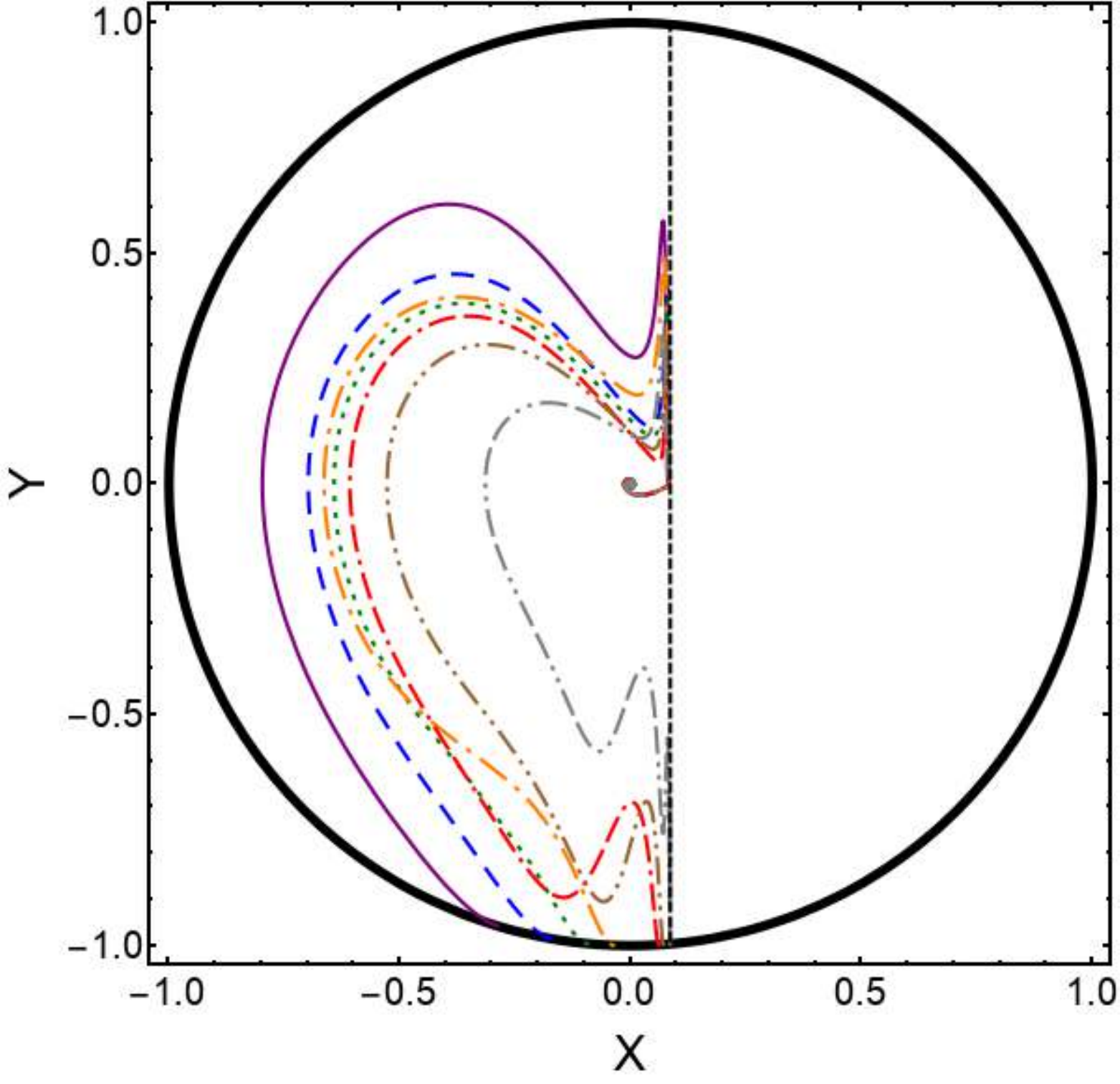} \ \ \ \ \ \ \ \ \ \
\includegraphics[scale=0.47]{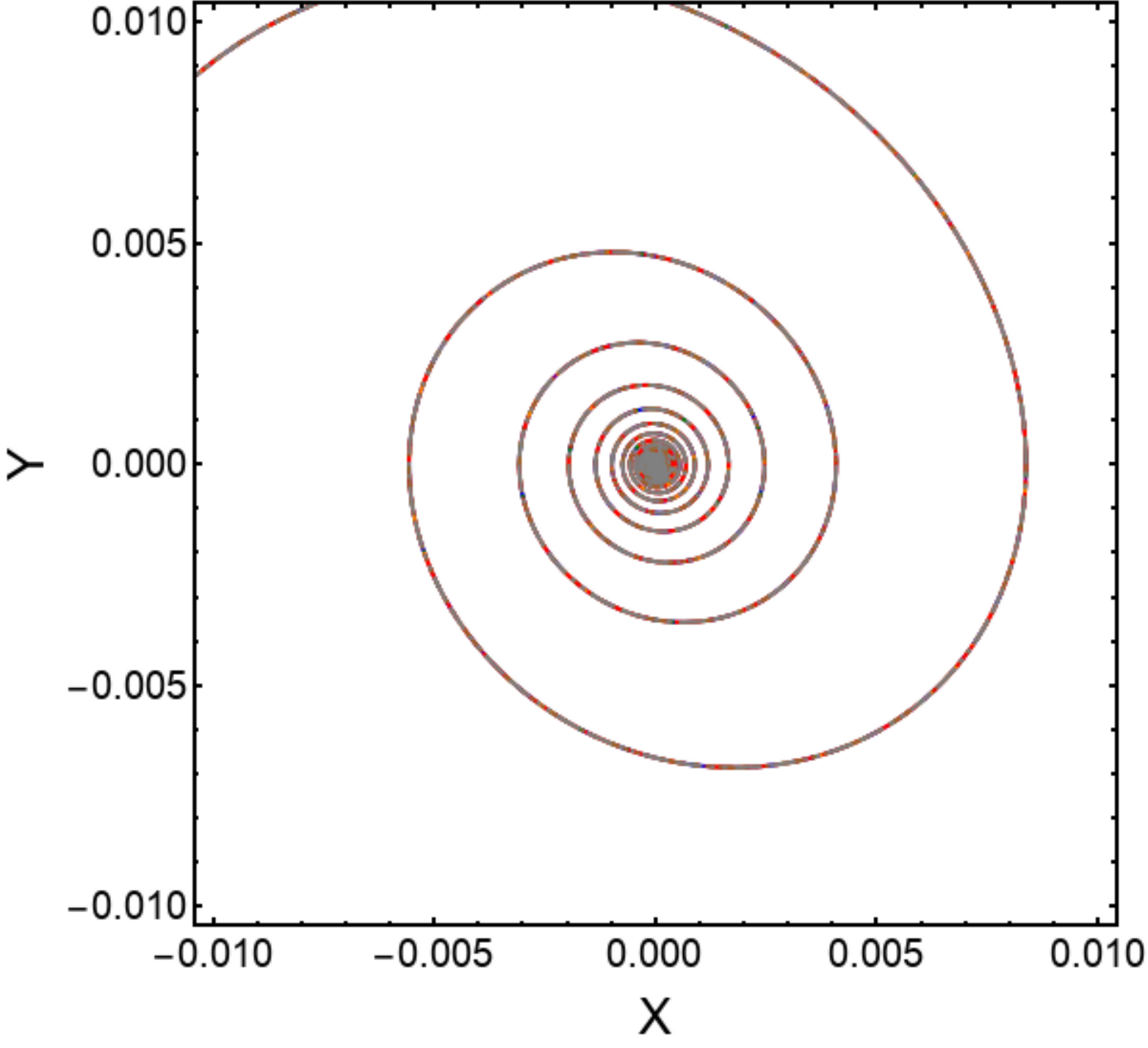}
\caption{Projection of 3 dimensional phase space portrait on $Z=0$ plane for linear dissipation coefficient with $m=0.62$, $C_{\mathrm{lin}} = 0.8$, $v_{0} = 35$, $\rho_{r0}= 10^{-3}$ and seven distinct initial conditions for $\phi_{0}$.}
\label{Phase-Space-Linear}
\end{figure}
\begin{figure}
\includegraphics[scale=0.3]{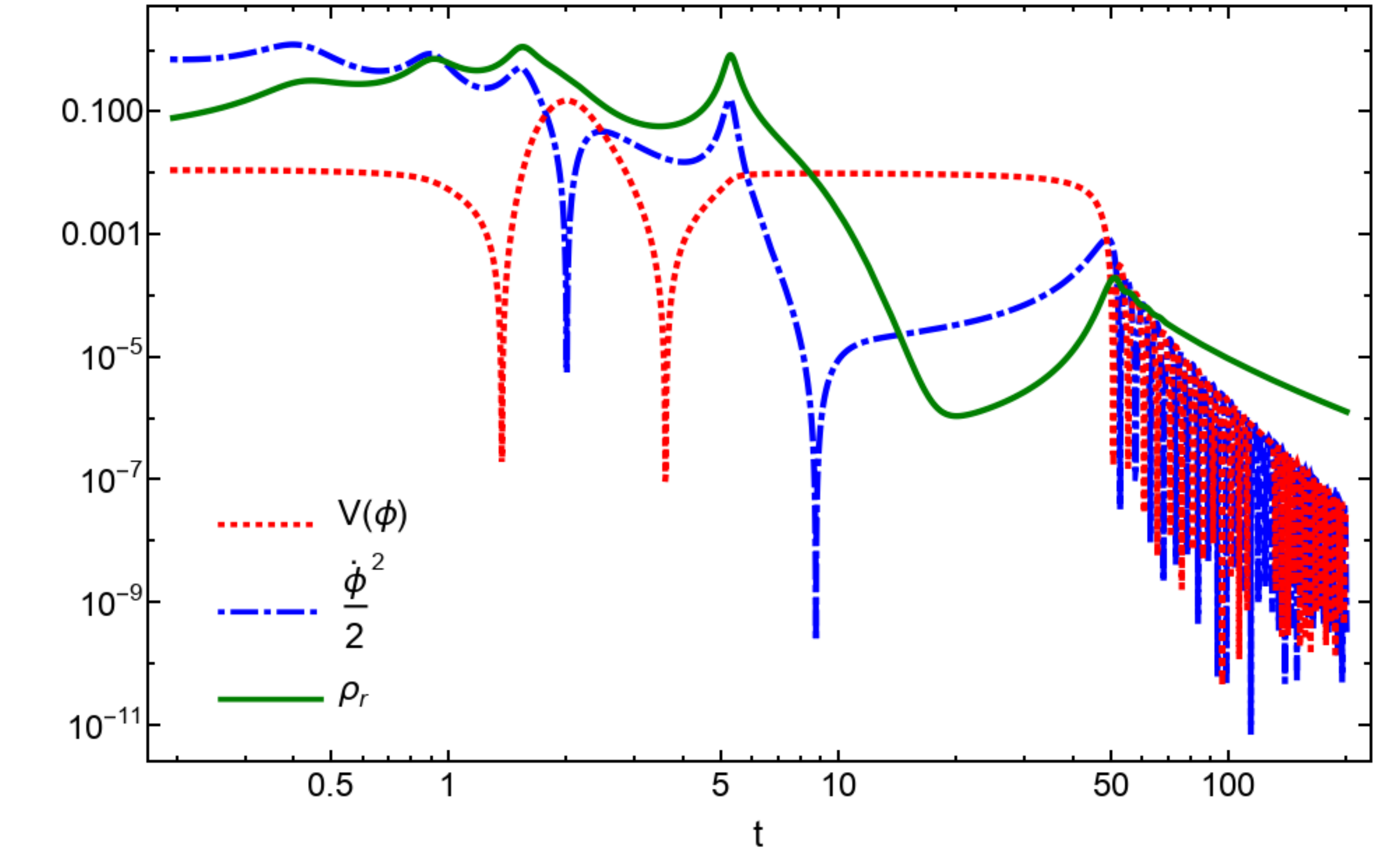}
\includegraphics[scale=0.36]{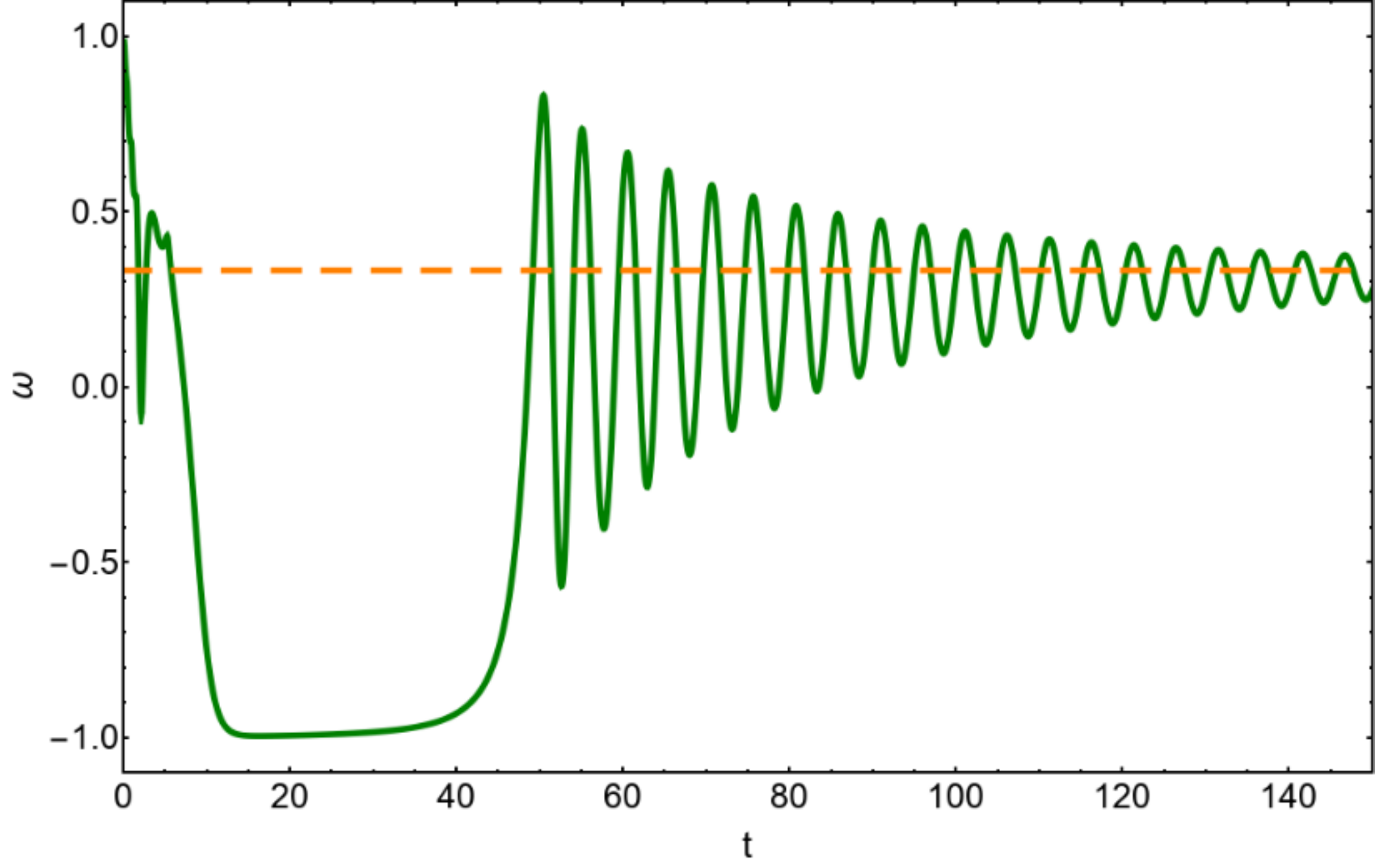}
\caption{Evolution of energy components as well as equation of state for linear dissipation coefficient with $m=0.62$, $C_{\mathrm{lin}} = 0.8$, $v_{0} = 35$, $\rho_{r0}= 10^{-3}$ and $\phi_{0} = 1.55$ (gray curve in Fig. \ref{Phase-Space-Linear}).}
\label{EWlinear}
\end{figure}

Fig. \ref{Phase-Space-Linear} illustrates the projection of the entire phase-space portrait on $Z=0$ plane for linear dissipation coefficient and $m=0.62$. The left plot shows that seven distinct initial conditions (different values of $\phi_{0}$ while other parameters are fixed) starting from the first bounce all exhibit an attractor behaviour and come to the center of circle (fixed point). In Fig. \ref{EWlinear}, the evolution of energy density components shows that inflation ends in radiation dominated epoch due to dissipative effects. Although such dissipative effects are large enough to sustain thermal bath during inflation, they are not large enough to suppress kinetic energy during inflation and terminate the universe in a radiation dominated epoch. Hence, there is very short kinetic dominated regime before the universe transits into radiation epoch which is typical in warm inflation scenario when the dissipation effect is small (such kinetic dominated regime after warm inflation has also been seen in \cite{Lima:2019yyv}). However, such a kinetic dominated regime does not have any adverse implications since it is very short. Moreover, since in this model such kinetic dominated regime  occurs around the minimum of potential, we see spiral behavior which is typical in cold inflation due to reheating phase. However, this oscillatory phase plays no role in making the universe hot and the universe enters into a radiation dominated regime due to radiation production during inflation and not a reheating phase. We also find that there is a radiation dominated regime before inflation phase as it was seen in case of cubic dissipation coefficient.

\begin{figure}
\includegraphics[scale=0.45]{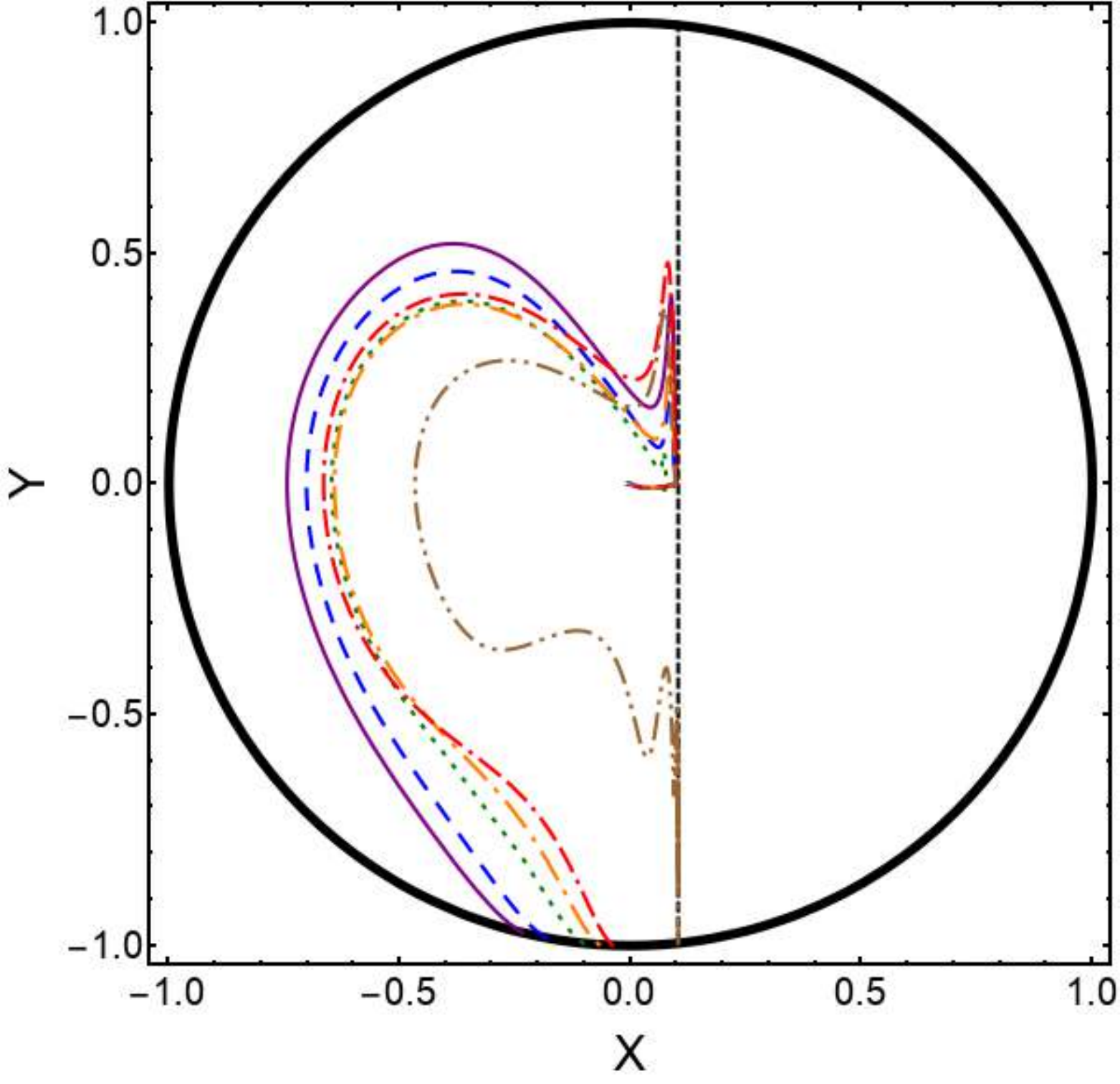} \ \ \ \ \ \ \ \ \ \
\includegraphics[scale=0.47]{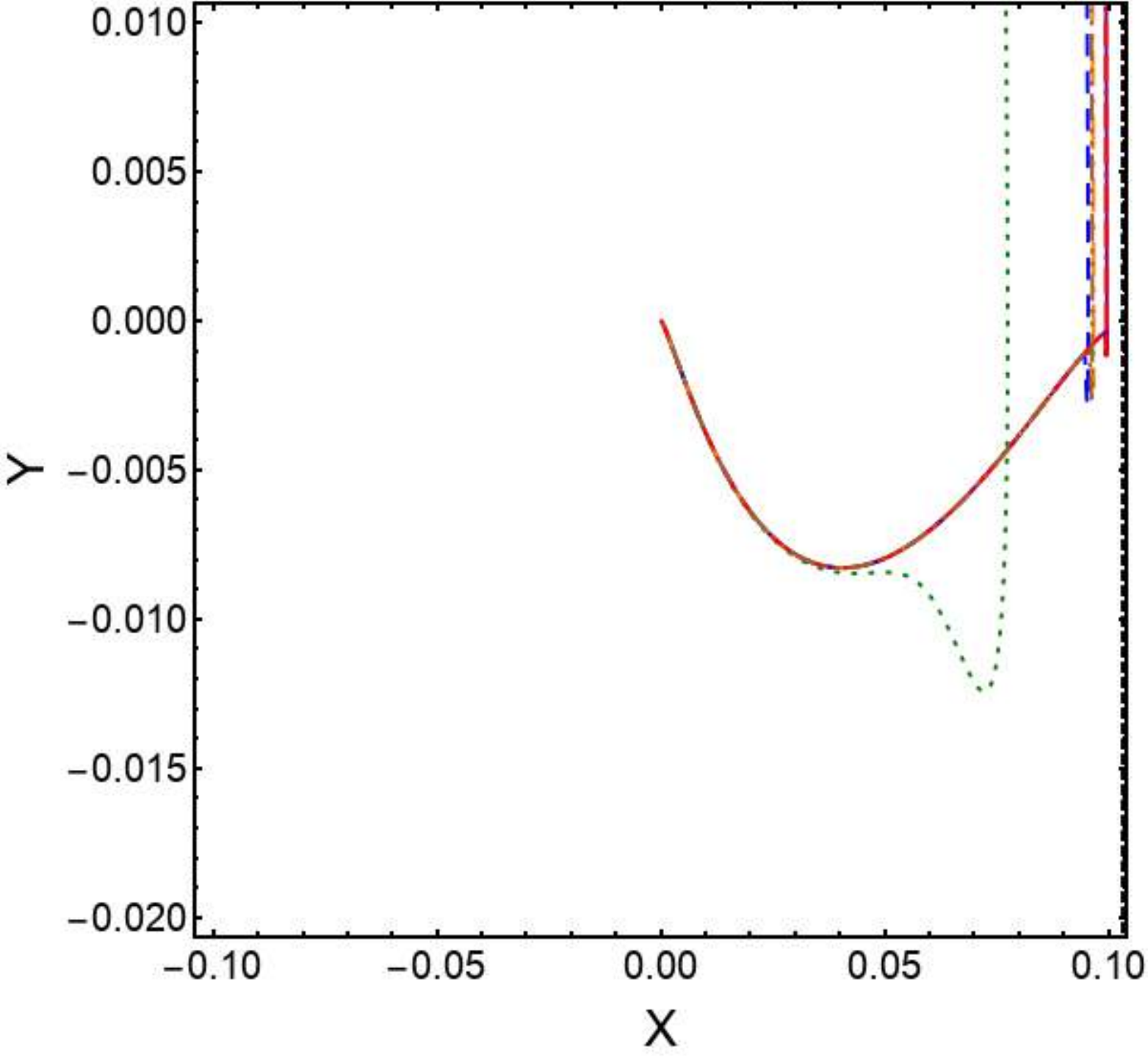}
\caption{Projection of 3 dimensional phase space portrait on $Z=0$ plane for inverse dissipation coefficient with $m=0.79$, $C_{\mathrm{inv}} = 0.2$, $v_{0} = 30$, $\rho_{r0}= 10^{-3}$ and six distinct initial conditions for $\phi_{0}$.}
\label{Phase-Space-inverse}
\end{figure}

\begin{figure}
\includegraphics[scale=0.3]{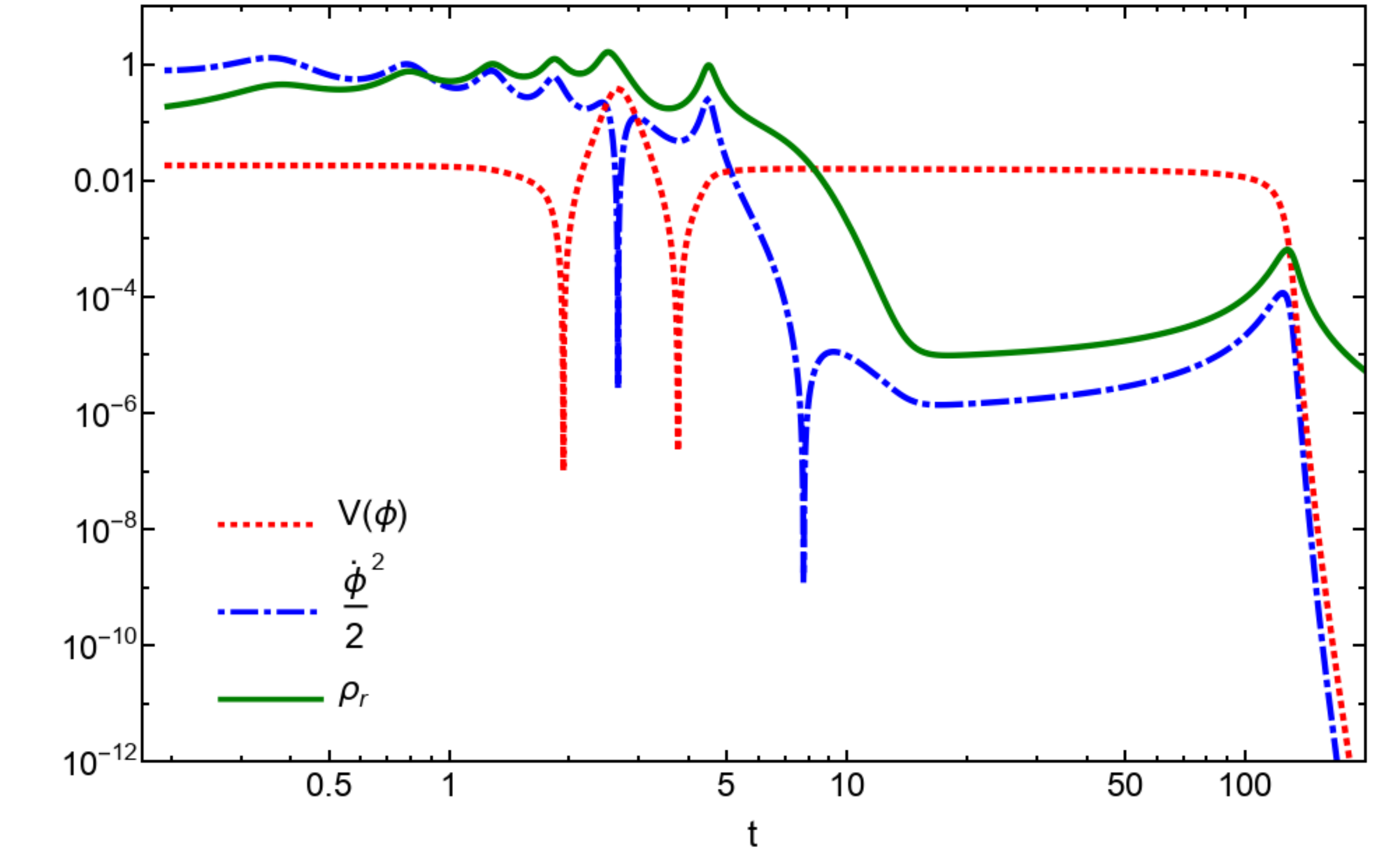}
\includegraphics[scale=0.36]{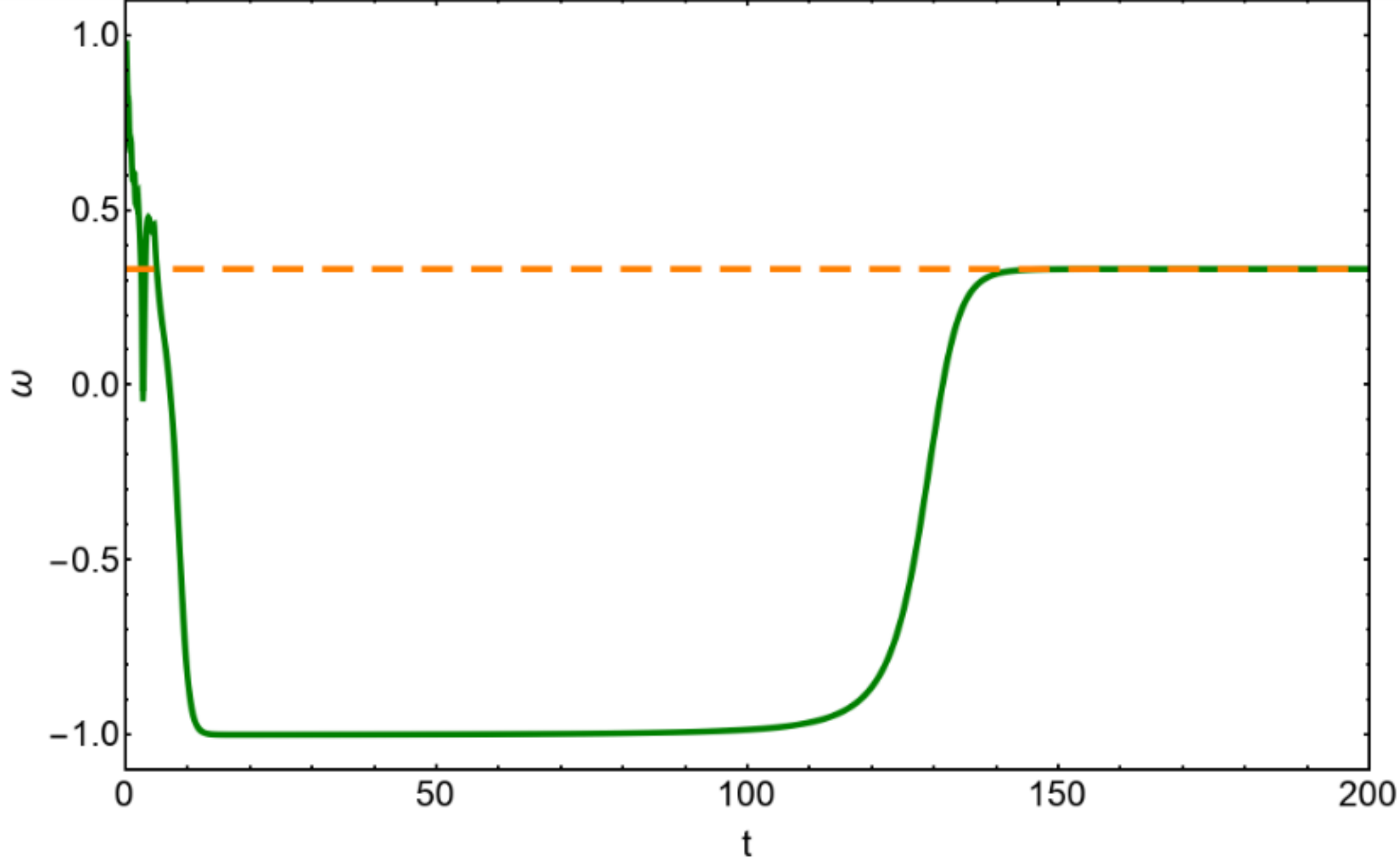}
\caption{Evolution of energy components as well as equation of state for inverse dissipation coefficient with $m=0.79$, $C_{\mathrm{inv}} = 0.2$, $v_{0} = 30$, $\rho_{r0}= 10^{-3}$ and $\phi_{0} = 2.2$ (brown curve in Fig. \ref{Phase-Space-inverse}).}
\label{EWinverse}
\end{figure}

In Fig. \ref{Phase-Space-inverse} we illustrate the projection of the entire phase-space portrait on $Z=0$ plane for inverse dissipation coefficient and $m=0.79$. We find that in this case although there is an attractor behaviour and all initial conditions starting from the first bounce come to the center of the phase space plot, there is no spiral behaviour. This is because  the universe enters into radiation dominated regime due to strong  dissipative effects. There is no oscillatory behaviour which is typical of cold inflation or what is found for previous cases of dissipation coefficient. In particular, the kinetic energy always remain sub-dominated during inflation due to strong dissipative effects and inflation ends when radiation becomes equal with potential energy. Furthermore, we find that there is also a radiation dominated regime before inflationary phase as it was seen for cubic and linear dissipation coefficients.

\section{Summary}

Onset of inflation in low energy inflationary models in classical theory is challenging in spatially closed models because of the recollapse of the universe before inflation can set in. Indeed one requies a high degree of fine tuning of initial conditions for the onset of inflation in such a case \cite{linde2003}. 
On the other hand if the spatial curvature is zero or negative this problem is non-existent and onset of inflation becomes highly probable \cite{compact, compact2, lindelecture}.
The recollapse in spatially closed model causes a big crunch singularity and a closed universe ends in a big crunch singularity in a few Planck seconds before the beginning of an inflationary phase. This is a long standing problem whose resolution becomes important given that current observational data favors low energy inflation models and a slight positive spatial curvature of the universe. There are various ways to overcome this issue in classical cosmology. Apart from the case of considering models with spatially flat and spatially open universe, one can consider a low energy inflation model coupled with an additional scalar field which drives an early phase of inflation near Planck density which is taken over later by low energy inflation. Further one can introduce additional higher order terms in quadratic potential to fit with Planck data \cite{lindelecture,linde2018}. But if one aims to understand this issue for a single field set up in low energy inflation models, 
 two issues need to be addressed simultaneously. The first is a successful and a generic resolution of singularities, and the second a mechanism to create favorable conditions for inflation to begin. The challenges underlying the first problem are well known and require insights from non-perturbative quantum gravity. The latter problem is  also non-trivial given that inflation in low energy inflation models begins at very small energy scales compared to Planck scale due to which initial conditions in the Planck regime are kinetic energy dominated which lead to a recollapse of the universe.  The above problem was recently analyzed in LQC \cite{gls20} where non-perturbative quantum gravity effects are known to result in a generic resolution of all strong curvature singularities \cite{generic}. In particular, the big bang/big crunch singularities are resolved and replaced by a non-singular bounce \cite{aps1,aps3,apsv}. It was found that for Starobinsky inflation potential, the universe in LQC cycles through various periods of expansion and contraction resulting in an onset of inflation even when inflaton starts from kinetic energy dominated initial conditions which result in a big crunch in a few Planck seconds.
At the heart of this resolution likes a hysteresis-like phenomena which changes the ratio of kinetic and potential energy in subsequent cycles in such a way that the equation of state even when starting from $\omega \approx 1$  becomes less than $-1/3$. The universe then enters a phase of accelerated expansion, a recollapse is avoided and inflation sets in.

Although hysteresis-like phenomena can enlarge the phase-space of initial conditions for plateau-like potential, with Starobinsky potential as the most known one, the inflationary phase either occurs after many cycles or does not happen due to flatness of potential for some unfavorable initial conditions \cite {gls20}. A pertinent question is whether there exists a mechanism which can result in onset of inflation in LQC even for such unfavorable initial conditions. The goal of the paper was to successfully answer this question. Motivated by Tolman's model in which the hysteresis-like phase happens due to entropy production sourced by viscous pressure and also warm inflationary dynamics, we considered spatially closed LQC in which the scalar field concurrently dissipates its kinetic energy into radiation field starting from the first bounce. Hence, there are two contributions for work done in each cycle; one from an asymmetric equation of state of scalar field during expansion-contraction phase and the other  from entropy production due to dissipative effects. Due to dissipative effects one expects the phenomena of hysteresis to become much stronger and inflation to set in far more easily.  %

We worked in the setting of effective spacetime description in LQC and obtained Hamilton's equations with non-perturbative quantum gravity corrections in presence of dissipative effects for $k=1$ model using holonomy quantization. The Hamilton's equations were numerically solved  for Starobinsky potential and three different dissipation coefficients inspired from 
quantum field theory all the way from the first bounce until the end of inflation. These were with cubic, linear and inverse temperature dependence. We found that even a small value of dissipation makes the hysteresis-like phenomena strong. The effect is such that inflation sets in not only a few cycles but also for those initial conditions which are extremely unfavorable for inflation to begin even in LQC without radiation production.
 Moreover, we find that as we make the dissipative effects large enough the hysteresis-like phenomena goes away and the universe inevitably enters into inflationary phase after just one bounce. To gain insights on the qualitative dynamics of the universe from the bounce until the end of inflation we studied the phase-space portraits for different initial conditions and all three dissipation coefficients. In all three cases, we found that all initial conditions experience an attractor dynamics showing that the universe goes through inflationary phase. In other words, the universe starting  from the first bounce with stiff-like initial conditions dilutes its kinetic energy due to both Hubble friction and dissipative effects and transfers it to radiation field, as opposed to dissipationless universe, whereby after some cycles the radiation energy density becomes the dominant energy component. However, such radiation dominated epoch continues only for a very short period since radiation energy density decays as $a^{-4}$. Then the potential energy becomes dominant resulting in inflationary phase. Moreover, if the dissipative effects are large enough, the inflationary phase will be of warm-type whereby the universe smoothly enters into radiation dominated epoch without the need for separate reheating epoch. Our analysis shows that with presence of dissipative effects, non-singular quantum gravitational dynamics results in an onset of inflation for low energy inflationary models even from highly unfavorable initial conditions. In comparison to the cases where dissipation is absent, we find hat inflation is set in much quicker due to stronger hysteresis-like phenomena. Since our results establish phenomenological viability of low energy inflationary models in spatially-closed universes at the level of background dynamics, it will be interesting to investigate the model at perturbative level to confront its predictions with observational data.

\section*{Acknowledgements}
This work is supported by the NSF grant PHY-1454832 and PHY-2110207. Authors thank Bao-Fei Li for discussions, and the anonymous referee for valueable comments.

\end{document}